\documentstyle[psfig,12pt]{aipproc}
\begin{document}

\title{ KALMAN FILTER BASED TRACKER STUDY FOR $\mu e$ - CONVERSION EXPERIMENT }

\author{Rashid M. Djilkibaev \thanks{Permanent address: Institute for Nuclear
 Research, 60-th Oct. pr. 7a, Moscow 117312, Russia}$^{,2}$, Rostislav V. Konoplich$^{2,3}$}
 \address{$^{2}$Department of Physics,
 New York University,
 New York, NY 10003 \\
 $^{3}$Manhattan College, Riverdale, New York, NY, 10471}
\maketitle

\begin{abstract}
The search for muon to electron conversion 
with a sensitivity of the order $10^{-17}$ requires a several order of magnitude
increase in muon intensity and a high resolution,
$\sigma \simeq$  0.1 MeV/c of the electron's momentum.

We present results of a pattern recognition  and track momentum
reconstruction algorithm that relies on a Kalman filter approach. 
Background from captured protons, neutrons, 
photons and from muon decay in orbit were generated by GEANT. 
The effective average 
straw tube background rate was 800 kHz.

The pattern recognition proceeds in two stages.
In the first, simple considerations using only  
straw tube center coordinates, without drift time information, were
applied to reduce the background to a manageable level. 
Then the drift time information is incorporated and a Deterministic Annealing Filter
applied to reach the final level of background suppression and 
to provide a starting 
point for the track momentum reconstruction using the Kalman filter.
This procedure reduces the simulated background by a factor 800
with small,(2.7\%), losses in real tracker hits.
 
The momentum resolution of the tracker is $\sigma $ = 0.12 MeV/c 
and the acceptance for muon conversion events with momentum
above 103.6 MeV/c is 22\%. These numbers do not differ significantly 
from the values obtained without background.

The expected number of events from muon decay in orbit (main background) 
in which the
decay electron has momentum greater than 100 MeV/c is 0.3, compared
to 6.5 $\mu e$ - conversion events above the same threshold 
for $R_{\mu e} = 10^{-16}$.

\end{abstract}

\section{Introduction}

The observation of $\mu e$ - conversion would provide the first
direct evidence for lepton flavor violation in charged lepton sector 
and require new physics,
beyond the Standard Model (see ~\cite{kuno} and references therein). 
Lepton flavor is not conserved in neutrino oscillations but the
modifications to the Standard Model to include the small neutrino
masses do not lead to an appreciable rate for $\mu e$ - conversion

\begin{center}
$\mu^{-} + N \to e^{-} + N$ .
\end{center}
 
This process violates the lepton flavor numbers, $L_{e}$ and $L_{\mu}$,
but conserves the total lepton number. The signature of the process is very
clear: a single monochromatic electron in the final state with the energy 
close to the muon mass:

\begin{center}
$E_{e} = m_{\mu} - B_{\mu} - E_{rec}$ 
\end{center}
where $m_{\mu}$ is a muon mass, $ B_{\mu}$ is a binding energy of the
1s muonic atom, $E_{rec}$ is a nuclear recoil energy. 

The SINDRUM II collaboration at PSI has carried out a program of experiments to 
search for $\mu e$ - conversion in various nuclei. They find that at 
90\% CL the upper limit for the reaction $\mu^{-} + Ti \to e^{-} + Ti$
is $6.1\times 10^{-13}$ ~\cite{psi}. In this experiment muons were accumulated
at a rate of $10^{7} \mu^{-}/sec$. According to preliminary results 
~\cite{psigold} for the reaction $\mu^{-} + Au \to e^{-} + Au$
a single event sensitivity is $3.3\times 10^{-13}$

In ~\cite{rashid} an idea of increasing of muon beam intensity
by a few orders of magnitude up to $10^{11} \mu^{-}/sec$ based on the 
solenoid-capture scheme was discussed and MELC experiment was proposed
~\cite{abadj} with a goal to reach a sensitivity of the order $10^{-17}$.
 
A new $\mu e$ - conversion experiment MECO (Muon Electron COnversion) E-940
~\cite{bnl}, exploiting the idea of the solenoid-capture scheme,
is under preparation at BNL. MECO aims to search for 
$\mu^{-} + Al \to e^{-} + Al$ with a single event sensitivity 2$\times$$10^{-17}$. It will
use a new high-intensity pulsed muon beam, which could yield about
$10^{11} \mu^{-}/sec$ stopped in a target. 

Also the PRIME (PRISM MuE conversion) working group at KEK expressed an
interest ~\cite{prime} to carry out a search for lepton flavor violation
in $\mu e$ - conversion in a muonic atom at a sensitivity of 
$10^{-18}$ using a proposed high intensity pure muon source of
$10^{11} - 10^{12} \mu^{-}/sec$ .  

In this article we develop a pattern recognition and  track reconstruction
procedure 
based on the Kalman filter technique for a transverse version of tracker for MECO
experiment. 

This paper is organized as follows: In section 2 the 
transverse tracker is described and the advantages of the tracker discussed.
In Section 3 possible backgrounds are discussed and  the 
procedure of background simulation briefly explained. 
In Section 4 the pattern recognition
procedure is developed. 
At this stage a deterministic annealing 
filter (DAF) is applied to make a final background 
suppression and provide a starting 
point for track momentum reconstruction by the Kalman 
filter technique. Section 5 describes the procedure of momentum 
reconstruction based on the Kalman filter. Results of the pattern 
recognition and momentum reconstruction are presented in this 
section. A brief summary outlook can be found in 
Section 6. The appendices provide a more detailed look at Kalman filter and deterministic
annealing filter. Also tracker resolution is discussed in Appendix.  

\section{The Transverse Tracker Description}

The goal of the MECO tracker is to detect the electron from $\mu e$ - conversion
with large acceptance and measure its momentum 
with high resolution ($\sigma  \simeq$  0.1 MeV/c).
The tracker is located in a uniform 1T magnetic field. 
The minimal tracker length is defined by the following
requirements:

Background electrons with an energy around 105 MeV  are produced
by cosmic rays  in the wall between the transport and detector
solenoids. These electrons cannot have a pitch angle $\theta$ in
the tracker greater than $45^{o}$, due to the adiabatic character
of charge particle movement. To suppress the  cosmic ray
background we require that measured pitch angle $\theta_{min}$ for
signal events to be more than $45^{o}$.

To get more redundancy for signal events we require that the measured
trajectory should have two full turns. This requirement sets
limits on the minimal tracker length, which is
expressed as:

\begin{center}
$ L_{Tracker}^{min} = 4 \cdot \pi \cdot 1/2.998 \cdot P \cdot cos(\theta_{min})/B \simeq
  4.19 \cdot 105 \cdot 0.707$/1.0 $\simeq$ 310 cm
\end{center}
where P is a muon conversion momentum (MeV/c) and B is a magnetic
field (1 Tesla) in the tracker region.

\begin{figure}[htb!]
\centerline{\hbox{\psfig{figure=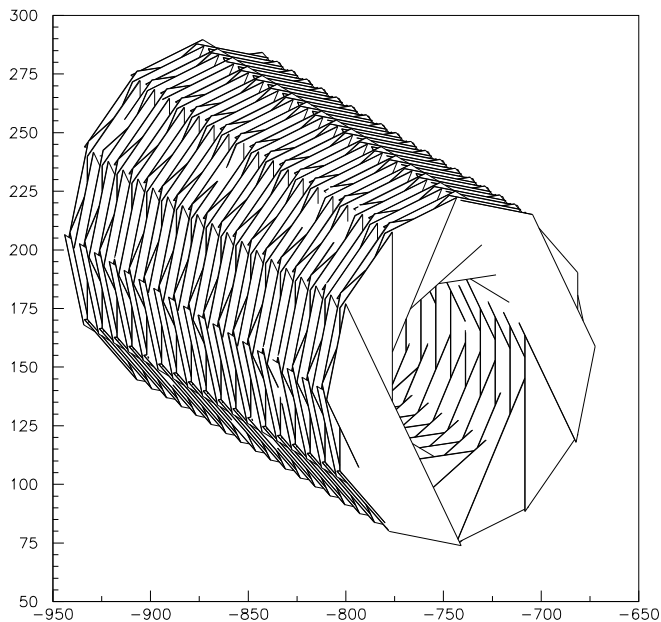,height=3.5in,clip=on}}}
  \caption{
  Schematic drawing of the tracker design.
 }
\label{fig:setup}
\end{figure}

The tracker consists of 18 modules spaced  17 cm apart. A module
consists of 6 planes, each turned at 30$^{o}$ and shifted 2.5 cm
relative to the previous one (Figure ~\ref{fig:setup}). A plane
consists of two trapezoidal chambers of  width 30 cm and lengths
70 to 130 cm in an up and down configuration. 
The chamber coordinate systems are defined by a
rotation  angle proportional to 30$^{o}$ giving an effective
``stereo'' of crossed directions for 12 different views.

A protective sheath is used to suppress background Compton
electrons from a chamber frame. 
The protective sheath makes nonsensitive a small
region of the anode wire near the frame.

The chamber consists of one layer of straw tubes (60 straws) of 5
mm diameter, and length varying from 70 cm to 130 cm. The total
number of chambers is 216. The chamber sensitive area starts from
38 cm radius. The straws are assumed to have wall 
thickness 15  $\mu$m and are
constructed of kapton. The total thickness of each chamber is 9
mg/$cm^2$. The total number of the tracker straws is 12960. The
tracker length is 302 cm. A signal from the straw anode wire 
will be used to get drift time.

The Al target is tapered in the downstream direction, with 5 cm
disk spacing and radii from 8.3 cm to 6.53 cm. The target is
placed in the graded portion of the DS magnetic field, with the
first disk at 1.75 T and the last at 1.3 T. Protons from the muon
capture are absorbed in three concentric polyethylene absorbers: a
conical tube of dimensions $R_{1}$ = 46 cm, $R_{2}$ = 70 cm, L =
260 cm and a tube (R = 70 cm, L = 200 cm) both have thickness 3
mm, and a tube of smaller radius (R = 36 cm, L = 235 cm) of
thickness 0.5 mm which are placed  before the tracker.

\bigskip
The transverse tracker is to be compared with a
longitudinal version that MECO is also considering. 
The longitudinal tracker ~\cite{bnl} consists 
of an octagonal array of eight detector planes placed symmetrically 
around the Detector Solenoid axis, plus eight more planes projecting 
radially outward from each vertex of the octagon. Each plane consists of
resistive straws approximately 300 cm long. A hit position 
in the radial and azimuthal direction is determined by the straw
position and the drift time on the anode wire. The hit position in 
the axial direction is determined by the centroid of the imaged charge 
from the anode wire, as collected on cathode pads.    

The transverse tracker would have several advantages in comparison with a
longitudinal one since: 

\noindent
$\bullet$ Normal non resistive straws are used without a pad system. \\
$\bullet$ Gas manifolds, straw end-caps and chamber supports are
all outside the conversion electron trajectories. \\
$\bullet$ Shorter straws (0.7 - 1.3 m) are technically easier to build
and are more robust against instability than longitudinal straws of 3 m.\\
$\bullet$ Complications from the small tilt of each plane of the longitudinal
tracker with respect to the magnetic field are avoided.\\
$\bullet$ Transverse geometry provides a simple signature of an event since
charged particles cross a single straw only once.\\
$\bullet$ There  is a significant simplification dealing 
only with single chamber hits points.\\
$\bullet$ The average number of spatially separated hits is a few times greater in  
comparison with the number of  spatially separated clusters in the longitudinal 
tracker. The hits are distributed uniformly in
lobes along the transverse tracker.\\
$\bullet$ The transverse tracker presents less material in the case of
15 $\mu$m straw wall thickness. The effective total thickness is
29*(15+2)*3.14 = 1550 $\mu$m versus longitudinal tracker 8*3.5*25*3.14 = 2200 $\mu$m.\\
$\bullet$ The pattern recognition  can be performed with good precision
even without drift time and amplitude information.\\
$\bullet$ The Cu layer covering straw tubes suppresses significantly gas diffusion
through a straw wall.\\
$\bullet$ Capacitive crosstalk  between channels is small.\\
$\bullet$ Low energy Compton electron backgrounds have distinguished signature.\\

\section{Background simulation}  

Our analysis is based on a full GEANT simulation taking into account an
individual straw structure.
Multiple scattering and energy loss are taken into account. 
Isobutane ($C_{4}H_{10}$) gas is assumed to fill the tubes. 

For the subsequent analysis we take only events satisfying the
following criteria: number of hits in the tracker is greater than 15;
MC simulated energy release in the calorimeter is greater than 80 MeV;
pitch angle is greater than 45 $^{0}$. That leaves about $35\%$ of
the original events.

The primary sources of charged particles in the tracker detector
during the detection time are protons, neutrons and photons from
muon capture by $^{27}Al$ nuclei and electrons from muon decay in
orbit.
The average and peak tracker rates from
different backgrounds are presented in Table ~~\ref{table:tab1}.
The main source of background to muon conversion is  muon decay in orbit (DIO).

\begin{center}
\begin{table}[htb!]
\caption { The Average and Peak (in parenthesis) Tracker  Rates }
\begin{tabular}{|l|c|c|c|c|c|}
\hline
Processes&proton&neutron &$\gamma$ &DIO & DIO\\
 & & & &$<$ 55 MeV &$>$ 55MeV\\
\hline
Particles/process & 0.1 & 1.2 & 1.8 & 0.9945 & $5.5 \cdot 10^{-3}$ \\
Particles/nsec & $120 (170)\times 0.1$ &$120 (170)\times 1.2$ &$120 (170) \times
1.8$ &80 (115) & 80 (115)\\
\hline
Prob. Particle to Hit Tracker&$ 1.08 \cdot 10^{-2}$&$0.92 \cdot 10^{-3}$
&$0.9 \cdot 10^{-3}$ & $4.0 \cdot 10^{-4}$ & $1.4 \cdot 10^{-2}$\\
\hline
Events in Tracker/30ns&3.9 (5.5) &4.0 (5.6) &5.9 (8.2) &1.0 (1.4) & 0.2 (0.3)\\
\hline
Straws on in Tracker/event &14.0 &5.2 &31.0 & 15.7 & 4.3\\
\hline
Straw Rate (kHz)&140 (200) &53 (75) &470 (650) &40 (56) & 2.2 (3.3)\\
\hline
\end{tabular}
\label{table:tab1}
\end{table}
\end{center}

To study the tracker performance in the presence of the background
the number of muon conversion events ($10^5$) with
initial momentum  of conversion electrons produced
in the target (105 MeV/c) was simulated and saved to a data file.
The expected number of
DIO events during the experiment time ($10^7$ sec) in
this region is $5.2 \cdot 10^4 $ events above 100 MeV/c. 

To study the main background 
the number of simulated  DIO events 
in this energy region  was chosen to be ten times greater.
Five different random backgrounds (protons, neutrons, photons and
DIO (see Table ~\ref{table:tab1}) were generated and saved to data
files. 

The  numbers in the table are the result of a GEANT simulations
at the anticipated beam intensity,
with no artificial increase as a  safety factor.
No suppression was added for rejection of
heavy ionization particles.

Background was taken from the data files and added to
conversion events and DIO events above 100 MeV/c. For example, we
simulate the number of events N for the proton background
according to a Poisson distribution with average 3.9 
taken from background Table ~\ref{table:tab1}. Then we randomly
pick-up N accidental proton events from the corresponding data
file and add these events to the muon conversion  or DIO event. 
In the
same way the above procedure is repeated for all background types.

\begin{figure}[htb!]
\centerline{\hbox{\psfig{figure=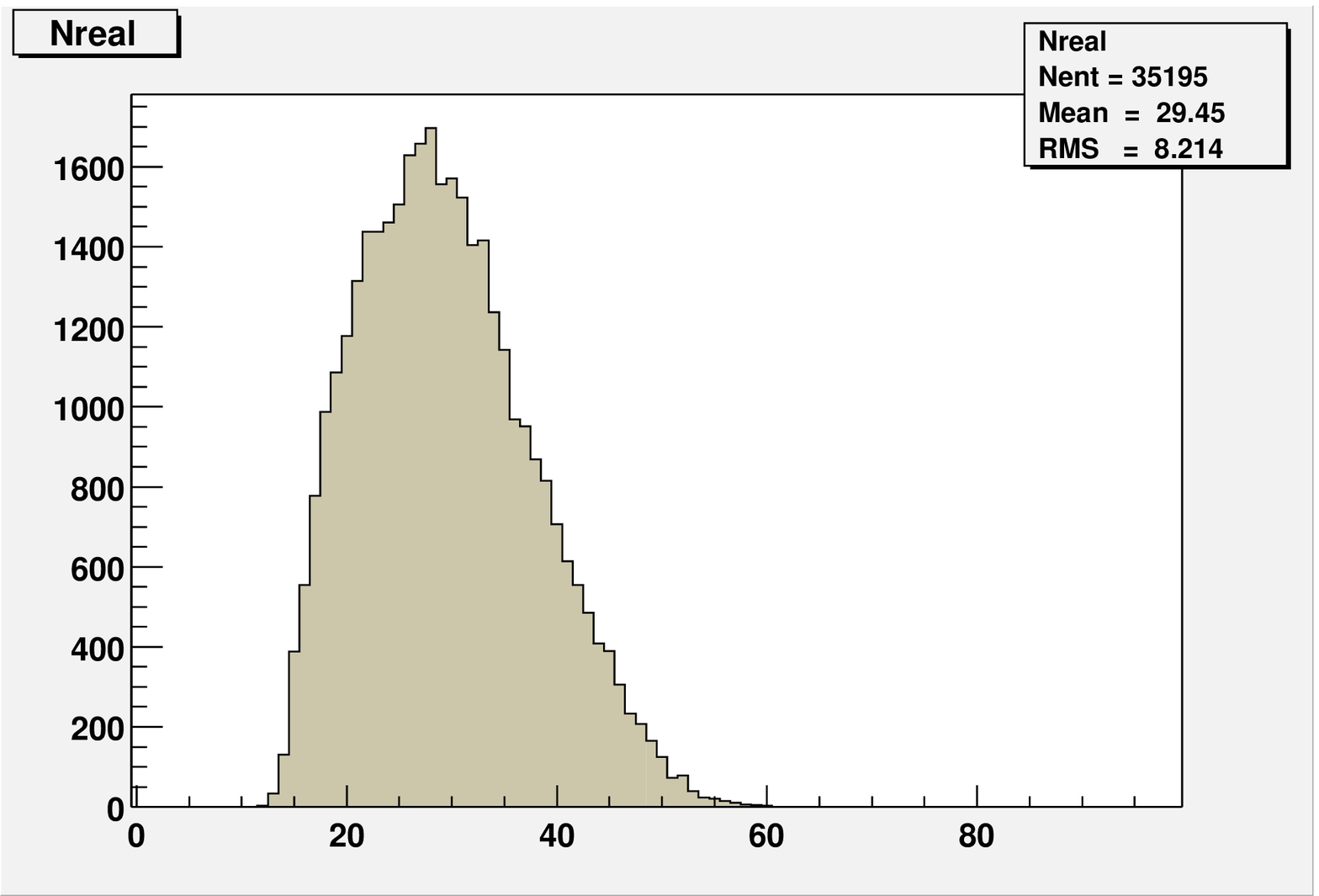,height=2.0in,clip=on}
                  \psfig{figure=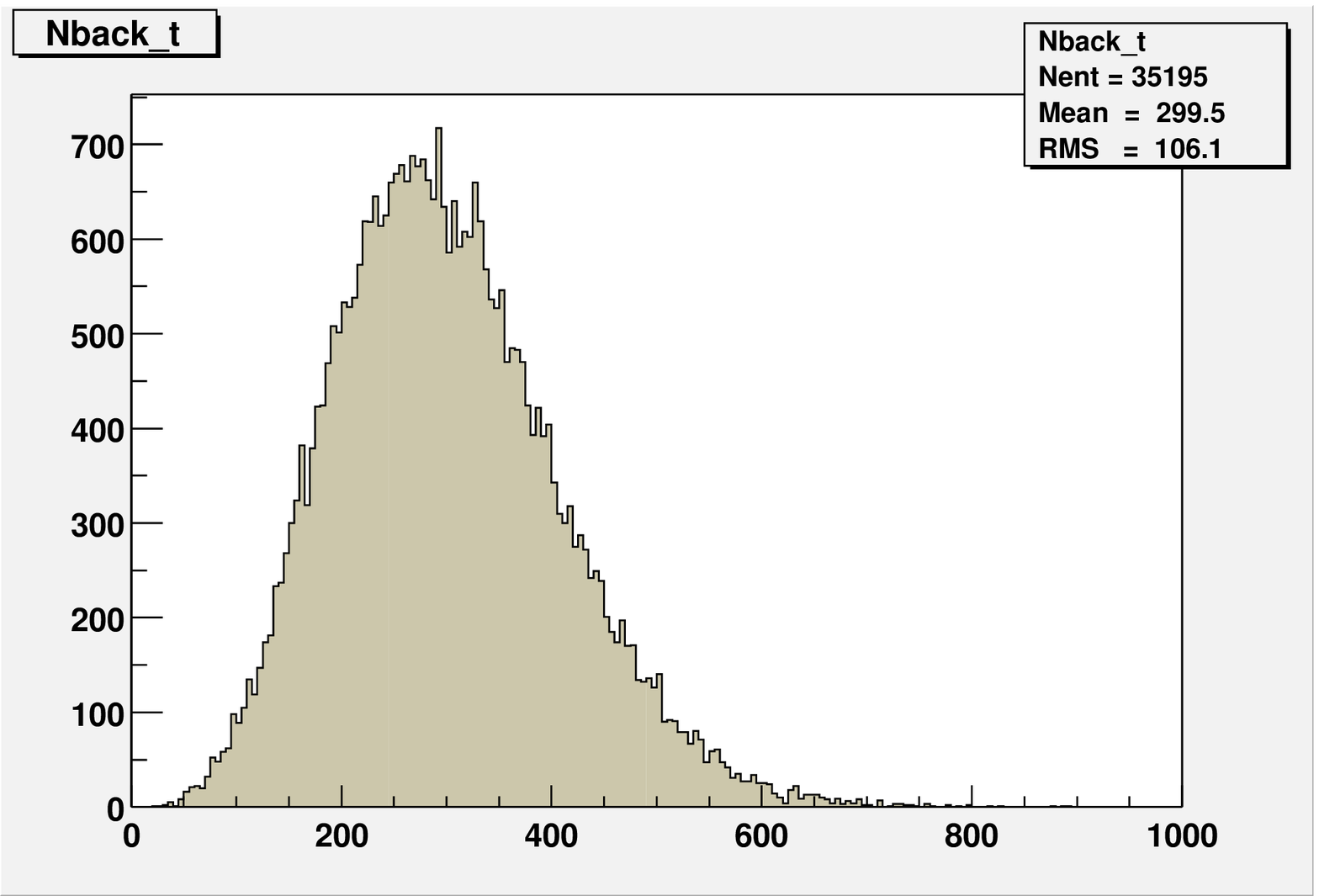,height=2.0in,clip=on}}}
 \caption{
Distributions in the number of real and background tracker hits. 
 }
\label{fig:nrb}
\end{figure}

In Figure ~\ref{fig:nrb} distributions in the number of real and background 
tracker hits are shown. 
The average number of real hits is 29.  
The real hits distribution starts from 15 hits, corresponding to 
the cut-off in the minimum number of selected hits. 
The maximum number of real tracker hits can reach 60.
The average number of background hits is 300 which corresponds to an average
straw rate about 800 kHz.
The distribution in the number of background hits is a broad one and
the maximum number of hits is about 900.

\section{Pattern recognition}

\begin{figure}[htb!]
\centerline{\hbox{\psfig{figure=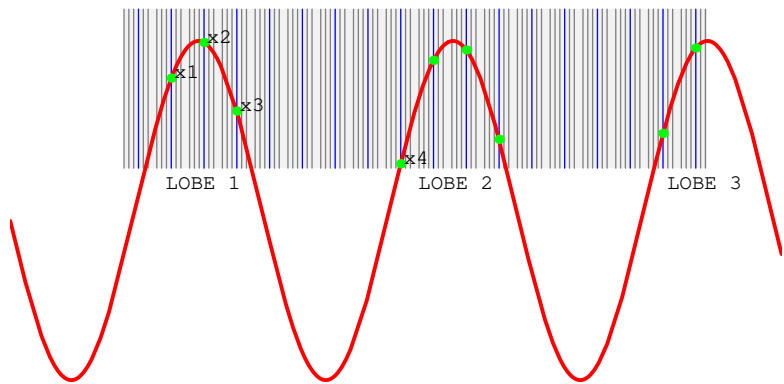,height=2.5in,clip=on}}}
  \caption{
  Lobes in the tracker.
 }
\label{fig:sin1}
\end{figure}

There are twelve views in the tracker,
each separated from the previous by 30 degrees. In a typical
event, a two-dimensional projection of the helical trajectory, a
sine curve, is observed in three or four of the views. There are
approximately 10 hits in each view, and 29 hits in the
event. The hits are grouped in lobes (see Figure ~\ref{fig:sin1}) with a typical
gap between lobes about 60 cm (in these gaps an electron travels in vacuum).
Therefore in the sensitive area of the tracker one gets track segments 
but not a complete track and the problem of combining information
from different lobes arises. Since in a given view at least 4 hits 
are required to reconstruct four parameters describing a 
two-dimensional projection of helix often it is not possible to get helix
parameters for a track segment in a single lobe and than to apply
track element merging strategy. Quite opposite one has to use
hits from different lobes for pattern recognition and then for
momentum reconstruction. We will show that an appropriate mathematical 
approach allows to resolve the problem..      

The sinusoidal projection of helix is described by four parameters:
$x_{0}^{\prime }$ , $z_{0}^{\prime }$, $R_{L}$, $R_{T}$,

\begin{equation}
x_{i}^{\prime }=x_{0}^{\prime }+R_{T}\cos (\frac{z_{i}-z_{0}^{\prime }}{R_{L}%
})
 \label{eq0}
\end{equation}

where importantly,  $R_{L}$ and $R_{T}$, related to the longitudinal and transversal
momenta, and z coordinate are common to all projections. Coordinate $x^{\prime}$ is 
defined in a system for each given view.

A simple and powerful reconstruction strategy
starts from a single chamber view. To determine the parameters in this view a minimum of four
hits are required.
The  total momentum and  momentum  pitch angle are determined.

To find the five parameters describing the full 3-dimensional helix,
 one additional hit outside the plane is required:

\begin{equation}
x_{i}^{\prime }=x_{0}^{\prime }+R_{T}\cos (\frac{z_{i}-z_{0}^{\prime }}{R_{L}%
}); ~~~~~~~
y_{i}^{\prime }=y_{0}^{\prime }+R_{T}\sin (\frac{z_{i}-z_{0}^{\prime }}{R_{L}%
})
 \label{eq01}
\end{equation}

The mathematical procedure to find the parameters is the following one:\\

\noindent
$\bullet$ four hits from a single view are taken to give the system of equations
for helix parameters in the coordinate system related with the view.\\
$\bullet$ the system of four equations is reduced to a unique equation for $R_{L}$. \\

\begin{equation}
\Delta x_{24}S_{43}S_{41}S_{31}+\Delta x_{14}S_{43}S_{42}S_{23}+\Delta
x_{34}S_{41}S_{42}S_{12}=0  \label{eq2}
\end{equation}

where $\Delta x_{ij}=x_{i}^{\prime }-x_{j}^{\prime }$ , $S_{ij}=\sin
((z_{i}-z_{j})/2R_{L}).$
\\
\noindent
$\bullet$ the equation is solved for $R_{L}$ numerically by the Newton-Raphson method.\\
$\bullet$ the parameters are expressed in terms of $R_{L}$. \\

\begin{equation}
\tan (z_{0}^{\prime }/R_{L})=\frac{\Delta x_{14}S_{43}\sin
((z_{4}+z_{3})/2R_{L})-\Delta x_{34}S_{41}\sin ((z_{4}+z_{1})/2R_{L})}{%
\Delta x_{14}S_{43}\cos ((z_{4}+z_{3})/2R_{L})-\Delta x_{34}S_{41}\cos
((z_{4}+z_{1})/2R_{L})}  \label{eq3}
\end{equation}

\[
R_{T}=\frac{\Delta x_{12}}{c_{1}-c_{2}}
\]

\[
x_{0}^{\prime }=\frac{x_{2}c_{1}-x_{1}c_{2}}{c_{1}-c_{2}}
\]

where $c=\cos ((z_{i}-z_{0}^{\prime })/R_{L}).$
\\
\noindent

$\bullet$ a fifth hit from a different view is included to
obtain the remaining parameters of helix. \\

\begin{equation}
x_{0}=\frac{1}{\sin (\alpha ^{\prime \prime }-\alpha ^{\prime })}\left\{
x_{0}^{\prime }\sin \alpha ^{\prime \prime }-[x_{5}^{\prime \prime
}-R_{T}\cos (\frac{z_{5}-z_{0}}{R_{L}}-\alpha ^{\prime \prime })]\sin
(\alpha ^{\prime })\right\}   \label{eq7}
\end{equation}

\begin{equation}
y_{0}=\frac{-1}{\sin (\alpha ^{\prime \prime }-\alpha ^{\prime })}\left\{
x_{0}^{\prime }\cos \alpha ^{\prime \prime }-[x_{5}^{\prime \prime
}-R_{T}\cos (\frac{z_{5}-z_{0}}{R_{L}}-\alpha ^{\prime \prime })]\cos
(\alpha ^{\prime })\right\} .  \label{eq8}
\end{equation}

where $\alpha ^{\prime }$ ($\alpha ^{\prime \prime }$) is the angle of
rotation from the global system to a given chamber system,

\begin{equation}
z_{0}=z_{0}^{\prime }-\alpha ^{\prime }R_{L}.  \label{eq6}
\end{equation}

It is important to note that the consideration of several lobes 
(see Figure ~\ref{fig:sin1}) allows calculation of the
longitudinal radius $R_{L}$ with high precision $\Delta P/P
\approx 10^{-4}$ due to the large distance between the lobes in
comparison with their size.

A two stage procedure was developed to provide the pattern
recognition in the transverse tracker:\\

$\bullet$ {\bf pattern recognition without drift time.} At this step only
information on centers of straw hits is used. The parameters of
the reconstructed average helix are considered as a starting point
to find an approximate helix
which fits the straw hit centers.\\

$\bullet$ {\bf pattern recognition with drift time.} A deterministic annealing 
filter DAF  ~\cite{daf} ~is applied to make a final background 
suppression and provide a starting 
point for track momentum reconstruction by the Kalman 
filter technique ~\cite{rkalm}, ~\cite{kalman}.

\subsection*{Pattern recognition  without drift time information}

In the first level  of the analysis procedure (the straw hit center
approximation) only centers of straw hits were used as chamber hit
coordinates $x^{\prime}$, z.
It was assumed that there is only one useful muon conversion 
or DIO track.

It is important to note that the straw hits allow us to
reconstruct the helix parameters with high precision even without
drift time information. Indeed, the uncertainty in the helix
radius $\Delta R_{T}$ is approximately  $\sigma \sim D/\sqrt12
\approx$ 0.14 cm for the diameter of a straw tube $D = $5 mm. The
average helix radius for muon conversion events is $R_T \sim $ 25
cm . We can therefore expect the momentum resolution $\Delta P/P 
\sim 0.5 - 1\%$
for a straw diameter of 5 mm.
It is worth noting that  using straw tube center positions
without drift time information we can reconstruct the total
momentum with the standard deviation $\sigma $ = 0.45 MeV/c for 
a straw diameter of 5 mm.

\begin{figure}[htb!]
\centerline{\hbox{\psfig{figure=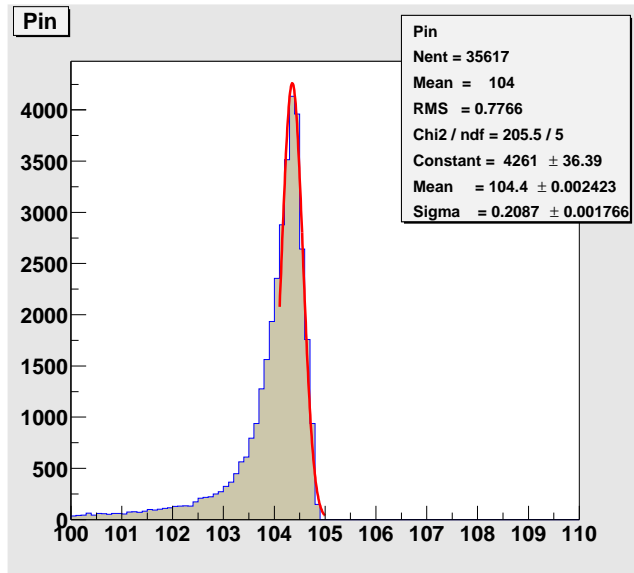,height=3.0in,clip=on}}}
 \caption{
 Momentum distribution of electrons entering the tracker.
 }
\label{fig:Pin}
\end{figure}

\bigskip

 The initial momentum distribution of electrons entering the
tracker and satisfying the criteria mentioned above is shown in
Figure ~\ref{fig:Pin}. This distribution is the a narrow one 
and  can be fitted by a Gaussian with a
standard deviation $\sigma $ = 0.21 MeV/c  and average momentum
P$_{in}$ = 104.4 MeV/c in the range 104.2 MeV/c - 105 MeV/c. 
The width of
the initial distribution of electrons is small in comparison with
the total width of the momentum distribution reconstructed in the
straw hit center approximation.

\begin{figure}[htb!]
\centerline{\hbox{\psfig{figure=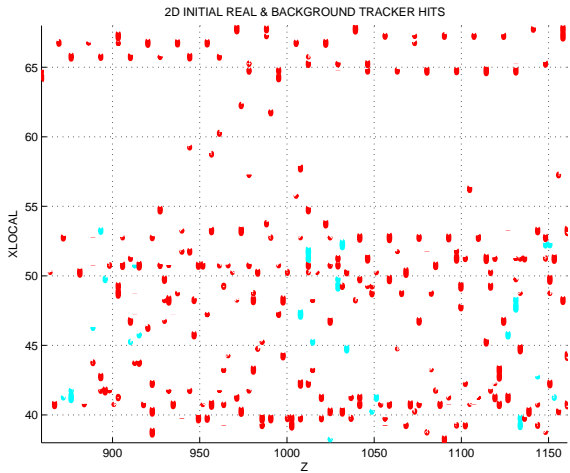,height=2.5in,clip=on}}}
 \caption{
 Plot of real + background tracker hits before the selection procedure.
 }
\label{fig:sel_0}
\end{figure}

In Figure  ~\ref{fig:sel_0} an initial map of real and background tracker 
hits in all tracker's views is shown as an example for a first simulated  event.
We will call it the sample event in the following.
The number of real tracker hits is 29 and the number of background 
hits is 260 in this event. The real hits are depicted as cyan spots 
and the background ones are shown as red spots. Note that in the 
figure all 12 views were joined in one for all 18 tracker modules. 

One can see from the initial distribution of tracker hits that 
some of background hits are concentrated along straight lines
( Figure  ~\ref{fig:sel_0}) providing a typical signature of Compton 
electrons from photon background.

 The procedure of the pattern recognition and
fitting without drift time in the presence of background 
can be described by the following steps:

\bigskip
1) {\bf Rejection of hits produced by Compton electrons.} \\

$\bullet$ 
A grouping containing 6 modules is taken for a given view. If in
this grouping there are $N_{1}>2$ hits with the same tube number
(same straw (x$^{\prime}$) coordinate) the corresponding hit numbers 
are recorded. 
If, in a second  grouping displaced by one module from the first, 
there are $N_{2}>2$ hits with the
same tube numbers as at the initial position of the grouping then 
hits from the first and the second grouping are marked as Compton electron hits. The 
procedure is repeated in forward and backward directions. At the
end of this step all marked hits are temporary removed from
the following analysis. This selection typically reduces twofold the number of 
background hits. 
Also about 5$\% $ of
the real hits are rejected by the procedure 
but later they can be restored.

\begin{figure}[htb!]
\centerline{\hbox{\psfig{figure=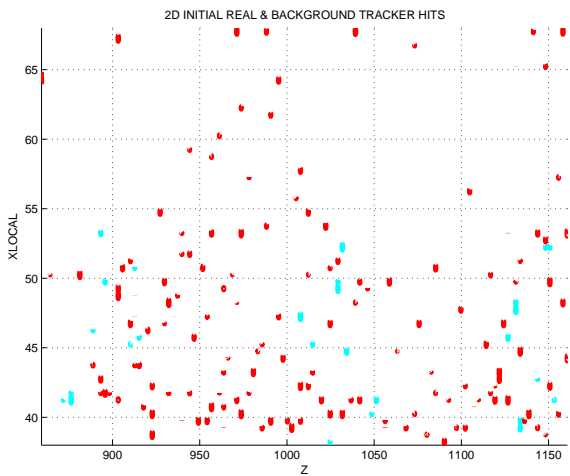,height=2.5in,clip=on}}}
 \caption{
 Plot of real + background tracker hits after the first selection procedure.
 }
\label{fig:sel_1}
\end{figure}

Figure  ~\ref{fig:sel_1} demonstrates the remaining tracker hits after application
of the rejection procedure to the sample event. The suppression of background hits 
is evident.    

\bigskip

2) {\bf Selection based on calorimeter and tracker hits.} \\

To reconstruct the helix parameters four hits are needed.
To reduce the number of four hit combinations 
the calorimeter hit is used in each combination. 
The  calorimeter coordinate resolution is $\sigma = 1.5$ cm which is
ten times greater than tracker hit resolution in the straw hit center
approximation. 
In spite of the poor coordinate calorimeter resolution 
it is possible to get a significant background suppression. 

This procedure is described by the following steps:

$\bullet$ For a given view all possible three hit
combinations are chosen and four hits are formed by adding 
a calorimeter hit to the three hit combinations.
The mathematical approach described above
is applied to four hits.
The algorithm allows us to calculate $R$ , $\cos \theta $ ,
$z_{0}^{\prime }$ , $x_{0}^{\prime }$ in the chamber system of
coordinates. Only the combinations that survive a cut-off in the
particle
momentum $p_{\min }=94MeV/c<p<p_{\max }=114MeV/c$ and pitch angle 
$\theta _{\min }=41.4^{\circ }<\theta <\theta _{\max }=66.4^{\circ }$ are
retained for subsequent analysis.

\bigskip
$\bullet$ A fifth hit from a different view than
the four hit combination is added. 
This allows us to calculate all 5 helix
parameters in the global system of coordinates. A cut-off in $x_{0}$ and 
$y_{0}$ is applied at this step to five hits.
The coordinates $x_{0}, y_{0}$ are required to be in an acceptable range
 $-40 < x_{0},y_{0} < 40 $ cm,
where the range is determined by the GEANT simulation

\bigskip
$\bullet$ We look for hits correlated in 3 dimensions to choose
good five hit combinations. For each selected five hit
combination using the found helix parameters we reconstruct the
helix, calculate all crossings of the tracker chambers for the
reconstructed trajectory and define the total number of crossings
N. Then we calculate how many times M at least one tracker hit
matches the crossing in a road $\pm $ 3 cm. In an
ideal case M should be equal to N.

\bigskip
$\bullet$ The probability is evaluated to get M hits in the  road. 
The probability that M  of N crossings  are in the road
is estimated by a trial function

\[
\mathit{Prob
}=\frac{N!}{M!(N-M)!}\epsilon^{M}(1-\epsilon)^{N-M}
\]

where ~~$\epsilon$ is 
the probability of that there is a hit 
in the road within $\pm$3 cm.

If the total probability \textit{Prob} is greater than the threshold
probability then the
given helix is considered as a good candidate.
The threshold probability and ~$\epsilon$ were found empirically
to be 0.001 and 0.95, respectively.

As an output of the previous steps we get a collection of valid
five hit combinations (helices). 
An individual hit is kept if it is in any valid hit
combination, providing a list of good tracker hits. It is worth
emphasizing that, due to the strong spatial correlations between the
helix hits in comparison with the un-correlated background, the
number of false hits is reduced drastically on applying the
road requirement. On average the number of background hits is 
reduced due to this step by a factor 20. 
 
\begin{figure}[htb!]
\centerline{\hbox{\psfig{figure=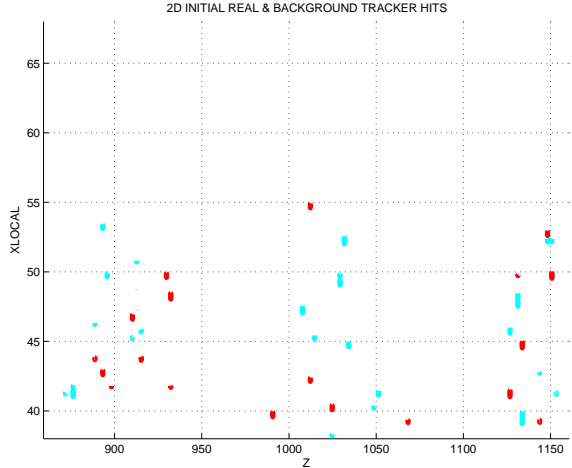,height=2.5in,clip=on}}}
 \caption{
 Plot of real + background tracker hits after the second selection procedure.
 }
\label{fig:sel_2}
\end{figure}

In Figure  ~\ref{fig:sel_2} 
the surviving hits are shown. 
The significant reduction of background hits can be seen by comparing 
 Figure  ~\ref{fig:sel_1} and  Figure  ~\ref{fig:sel_2}.

Note that at this step no real hits are removed for the sample event. 

\bigskip

3) {\bf Selection based on tracker hits.}\\

In this selection procedure the tracker hits found in step 2 are used.
This procedure repeats the previous one with the following changes:\\

$\bullet$ Hits are chosen only from the tracker hits which allows to
improve the performance of the selection procedure since tracker hit 
coordinates are defined significantly better than calorimeter hit 
coordinates.

\begin{figure}[htb!]
\centerline{\hbox{\psfig{figure=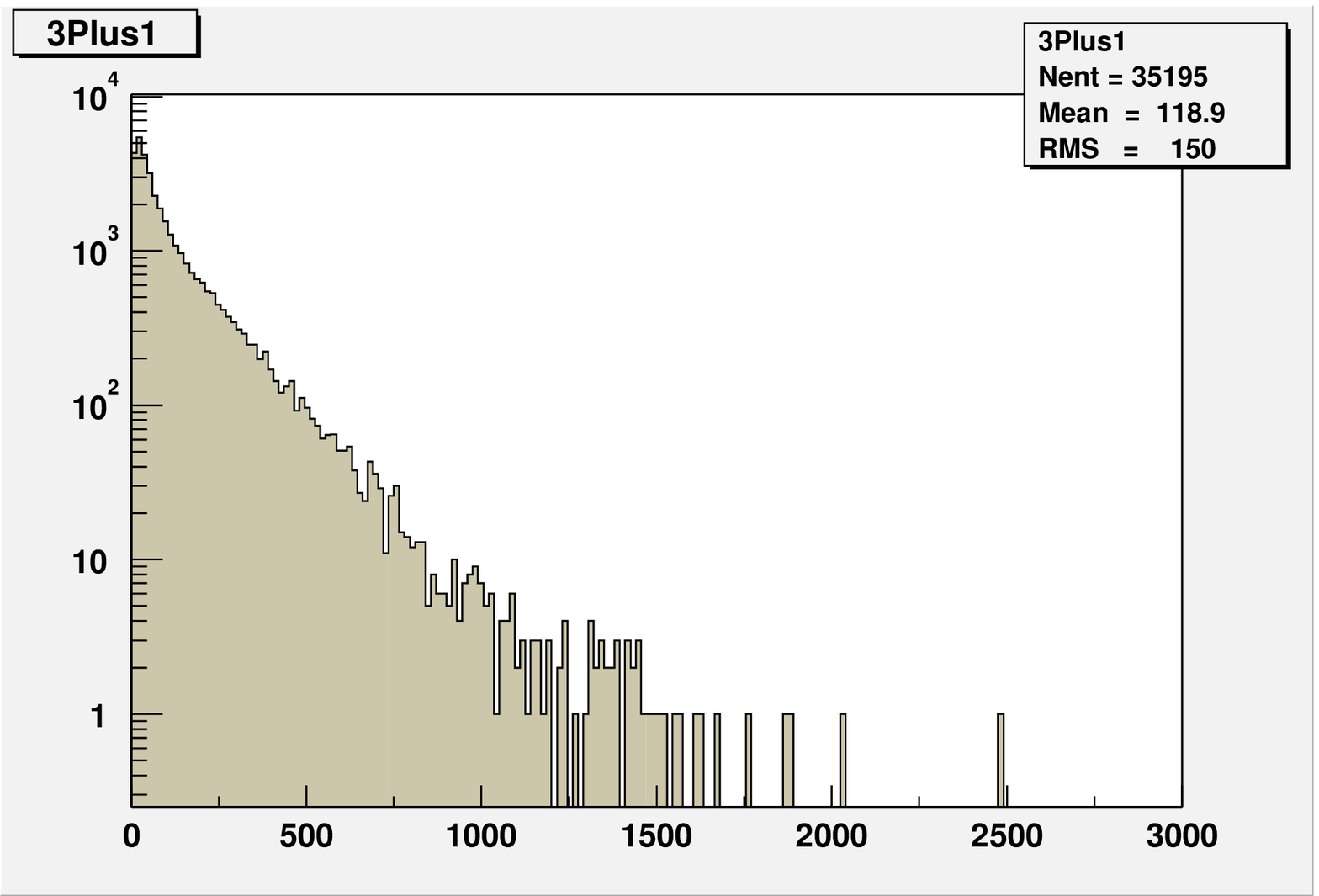,height=2.0in,clip=on}
                  \psfig{figure=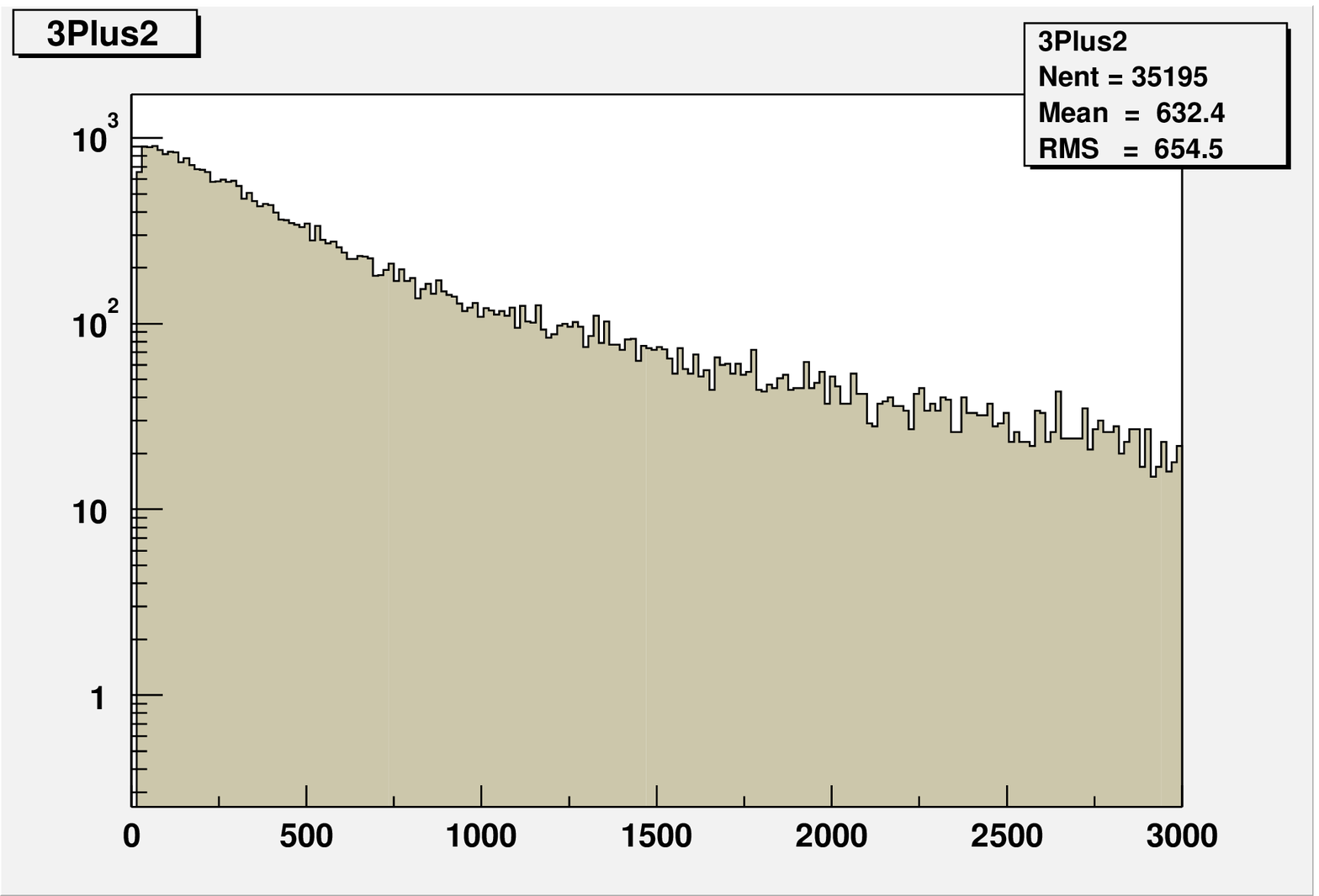,height=2.0in,clip=on}}}
 \caption{
Number of four hit and five hit combinations.
 }
\label{fig:comb}
\end{figure}

A plot of the number of four hit and five hit
combinations in the presence of the background (in average about 120 
and 630 respectively) is shown in Figure ~\ref{fig:comb} (a) and (b).

As seen in Figure ~\ref{fig:comb} these distributions are
broad ones reaching a few thousand combinations.

\bigskip
$\bullet$ For each five hit combination selected above we evaluate
the position of a hit in the
calorimeter $x_{eval}$ , z$_{eval}$ on the basis of the defined helix parameters.
The hit combination is accepted
only if the evaluated hit matches the calorimeter hit within a road : $%
\left| x_{calo}-x_{eval}\right| < 7 cm$ .

$\bullet$ Parameters at this step are more
restrictive: the road width is taken to be 0.75 cm and the
probability to be within the road $\epsilon$ = 0.97. 

\begin{figure}[htb!]
\centerline{\hbox{\psfig{figure=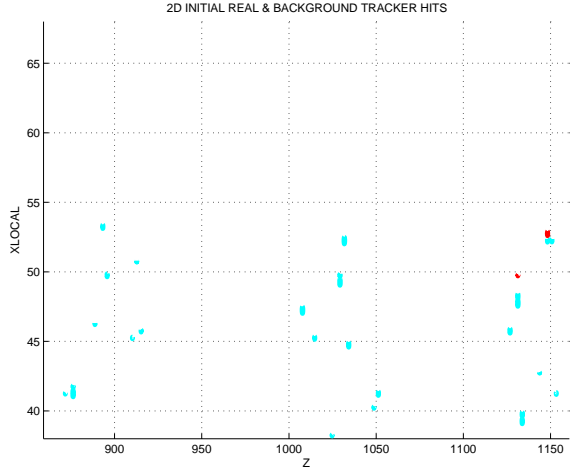,height=2.5in,clip=on}}}
 \caption{
 Plot of real + background tracker hits after the third selection procedure.
 }
\label{fig:sel_3}
\end{figure}

 Figure  ~\ref{fig:sel_3} shows tracker hits 
remaining after this step for the  sample event. 
The number of background hits is 
reduced typically by a factor 4 - 5 due to this step.

\bigskip
4) {\bf Selection including restored tracker hits.} \\

$\bullet$ Some real tracker hits are lost
at previous steps of the procedure especially at Compton electron hits
rejection. To restore the lost hits an average helix for a given event is
reconstructed and tracker hits that match the average helix in a road
$\pm$1 straw are added to the list of valid tracker hits obtained above.

$\bullet$ Step 3 of the procedure is repeated for the extended list of tracker 
hits. On average the list of valid
hits is extended by 1 real and 2 background hits.

$\bullet$ The union of all hits in valid five hit combinations 
in the extended list is
used to provide an input for an average helix.
A global fit is applied to reconstruct the helix parameters on the basis of
the list of the selected tracker hits and the parameters of the average helix are
considered as a starting point for the fit.  \\ 

In this section the pattern recognition procedure without drift time
based on the straw hit center approximation was developed.
In this approximation no ambiguity due to mirror hits arises 
and the total momentum is reconstructed  
with the standard deviation $\sigma $ = 0.45 MeV/c. 
The  selection procedure based on 3D space correlations between real
tracker hits inside the road significantly reduces background.
The overall background rejection factor is about 130
for the pattern recognition procedure without drift time.

\subsection*{Pattern recognition with drift time }

\begin{figure}[htb!]
\centerline{\hbox{\psfig{figure=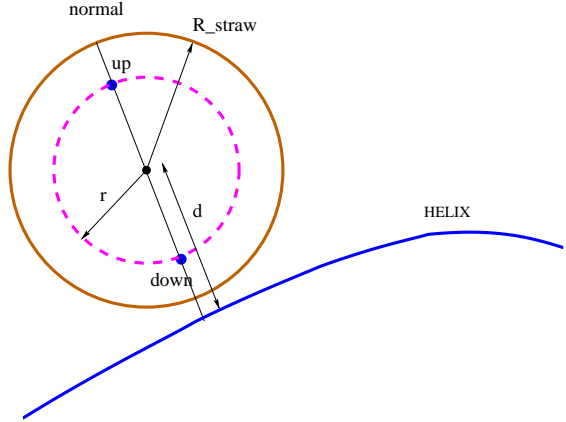,height=2.2in,clip=on}}}
 \caption{
 Reconstruction of up and down points by the helix obtained in the
 straw hit center approximation.
 }
\label{fig:TTRS1}
\end{figure}

The second stage in the pattern recognition procedure uses the
fitted helix and drift time for hits selected at the first stage.
The reconstruction of helix parameters can be improved by taking
into account that in addition to a straw coordinate for
each hit a chamber gives the radius r calculated from the measured drift
time $t_{i}^{meas}$. The errors ($\sigma $) in radius measurements
were taken to be 0.2 mm. This radius r carries an ambiguity as to
whether the track passed left or right of the wire. The search of
two possible up and down hit positions lying on the circle of
radius r is based on the fitted helix obtained previously in the straw
hit center approximation. 
We can call these up and down points as true and mirror ones.
Up and down points are extracted (see
Figure ~\ref{fig:TTRS1}) from the intersections of a normal to the
helix through the straw center and the circle of the drift radius
r. In this case coordinates of up and down points are given by

\begin{equation}
x_{\prime} = x_{c} \mp r sin \beta ;
~~~~~z = z_{c} \pm r cos \beta ;\\
~~~~~tan\beta = -(R_{T}/R_{L})sin((z_c-z_{0}^{\prime})/R_{L})
\label{equpdown}
\end{equation}

\begin{figure}[htb!]
\centerline{\hbox{\psfig{figure=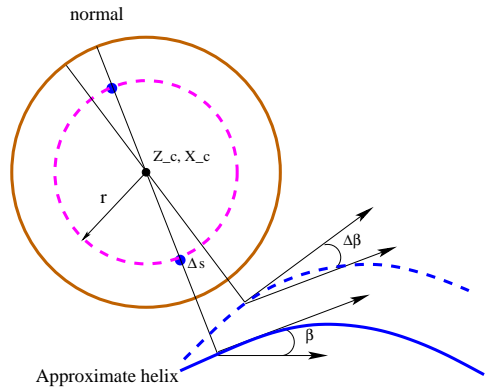,height=2.0in,clip=on}}}
 \caption{
Uncertainties in up and down point positions.
 }
\label{fig:TTRS2}
\end{figure}

The uncertainties $\Delta s$ (see Figure ~\ref{fig:TTRS2}) 
in the determination of up and down point
positions are
small and can be evaluated in the following way.
The direction of 2D helix in the chamber coordinate system is given by
$ tan\beta $ (see Figure ~\ref{fig:TTRS2} and Eq.(\ref{equpdown})).

Since the uncertainty in $R_{T}$ is the dominant one the uncertainty in the
direction of the helix is of the order 
$\Delta \beta /\beta \sim \Delta
R_{T}/R_{T} \sim \Delta
R/R \sim \Delta P/P$  where $\Delta P/P<0.01, \beta \sim 1$.

The uncertainty in the determination of up or down point
position \\
$\Delta s \sim \Delta \beta $ r  $ <
\beta (\Delta P/P)r_{tube} \sim 1 \times 10^{-2} \times 0.25$ cm $ \sim 25 \mu$m
which is much less than a measurement precision 200 $\mu $m.

To reject background hits remaining after pre-selection  stage and
to resolve the up - down ambiguity
the deterministic annealing filter (DAF) and the Kalman filter (KF) 
are applied.

In principle at this stage the Kalman filter (KF) approach \cite{kalman}
could be applied. However an application of the Kalman filter 
(see Appendix A) requires that the problem of assignment of hits to a track
has been entirely resolved by the preceding selection procedure. 
If this is not the case the filter has to run on every possible 
assignment choosing the best one according to the chi-square 
criterion. For the number of tracker hits greater than 15 this 
combinatorial search is computationally expensive and practically
unfeasible. Therefore as the last step of pattern recognition 
we will use the deterministic annealing filter (DAF) ~\cite{daf}.

DAF is a Kalman filter with re-weighted observations (see Appendix B).
For the DAF procedure we introduce artificial layers placed at the chamber straw centers 
in order to have competing true and mirror points in one  layer. 
To overcome the problem of insufficient information in the initial
phase of the filter, an iterative procedure is applied. After a first
pass of filter and after smoothing, the track position can be predicted 
in every layer of the tracker. Based on these predictions, the assignment
probabilities for all competing hits can be calculated in every layer.
If the probability falls below a certain threshold, the hit is excluded 
from the following consideration. The assignment probabilities of
the remaining hits are normalized to one and used as the weights
in the next iterations of the filter.

In our case the operation of DAF is described by the following steps:\\

$\bullet$ every true and mirror  point is projected on a 
layer corresponding to the center of straw in the direction
defined by the fitted helix for a given event.\\

$\bullet$ initial probabilities of competing points
in the layer are assumed to be equal.\\

$\bullet$ annealing schedule is chosen according to the following
formula\\
  $V_{n}=V (\frac{50}{f^{n}}+1)$ for a variance of observations,
where V = $\sigma^{2}$ and $\sigma$ = 200 $\mu$m. 
The annealing factor f is chosen to be either 1.4 or 2.\\

$\bullet$ standard Kalman filter runs on all layers taking 
observations as weighted mean according to assignment probabilities.\\

$\bullet$ the filter runs in the opposite direction, using the
same weighted mean as the forward filter. By taking a weighted mean of
the predictions of both filters at every layer, a smoothed state vector
and its covariance matrix are obtained.\\ 

$\bullet$ based on these predictions and the  covariance matrix, the
assignment probabilities of the hits are calculated. If combined 
hit probability for true and mirror point  falls below a 
certain threshold ($10^{-7}$), the hit is rejected. The assignment probabilities of
the remaining points in the layer are normalized 
to one and used as the weights
in the next iterations of the filter.\\ 

$\bullet$ iterations in n stop if one of the following
conditions is satisfied:\\
1)for the reconstructed track $\chi^2 > \chi^2_{max}$ 
where $\chi^2_{max}$ = 1000 or $\chi^2 < 0$.\\
2)for the reconstructed initial momentum $P_{in} < 94 MeV/c$ or
$P_{in} > 114 MeV/c$\\
3)variation of $\chi^2$ is small in comparison with the 
previous iteration $|\chi^2_{n+1}-\chi^2_{n}|/\chi^2_{n}<0.01$.\\   

$\bullet$ when iterations stop (on average after 7 iterations) 
The DAF procedure is repeated for the different annealing factor f.
From results corresponding to two annealing factors we choose 
that one corresponding to the minimum $\chi^2$.\\

\begin{figure}[htb!]
\centerline{\hbox{\psfig{figure=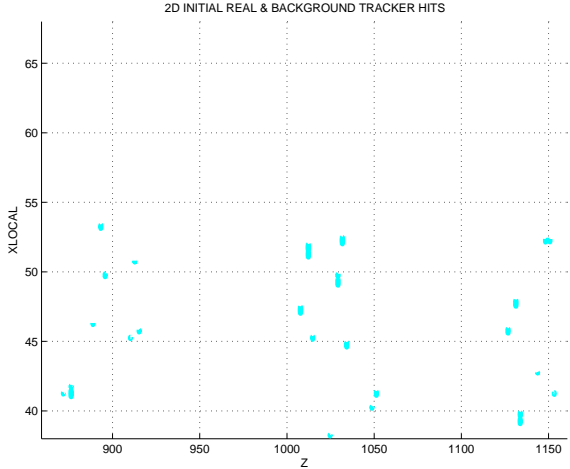,height=2.5in,clip=on}}}
 \caption{
 Plot of real + background tracker hits after the DAF selection procedure.
 }
\label{fig:sel_5}
\end{figure}

Figure  ~\ref{fig:sel_5} shows the distribution of tracker hits for the sample event
(recall that initially we had 260 background hits for this event).  
There are no surviving background hits and missing real hits
in this case. Reconstructed lobes are clearly seen in the 
Figure. We conclude that DAF is effective in rejection
of background hits remaining after pre-selection procedure and 
also it provides a good starting point for the track reconstruction.

\begin{figure}[htb!]
\centerline{\hbox{\psfig{figure=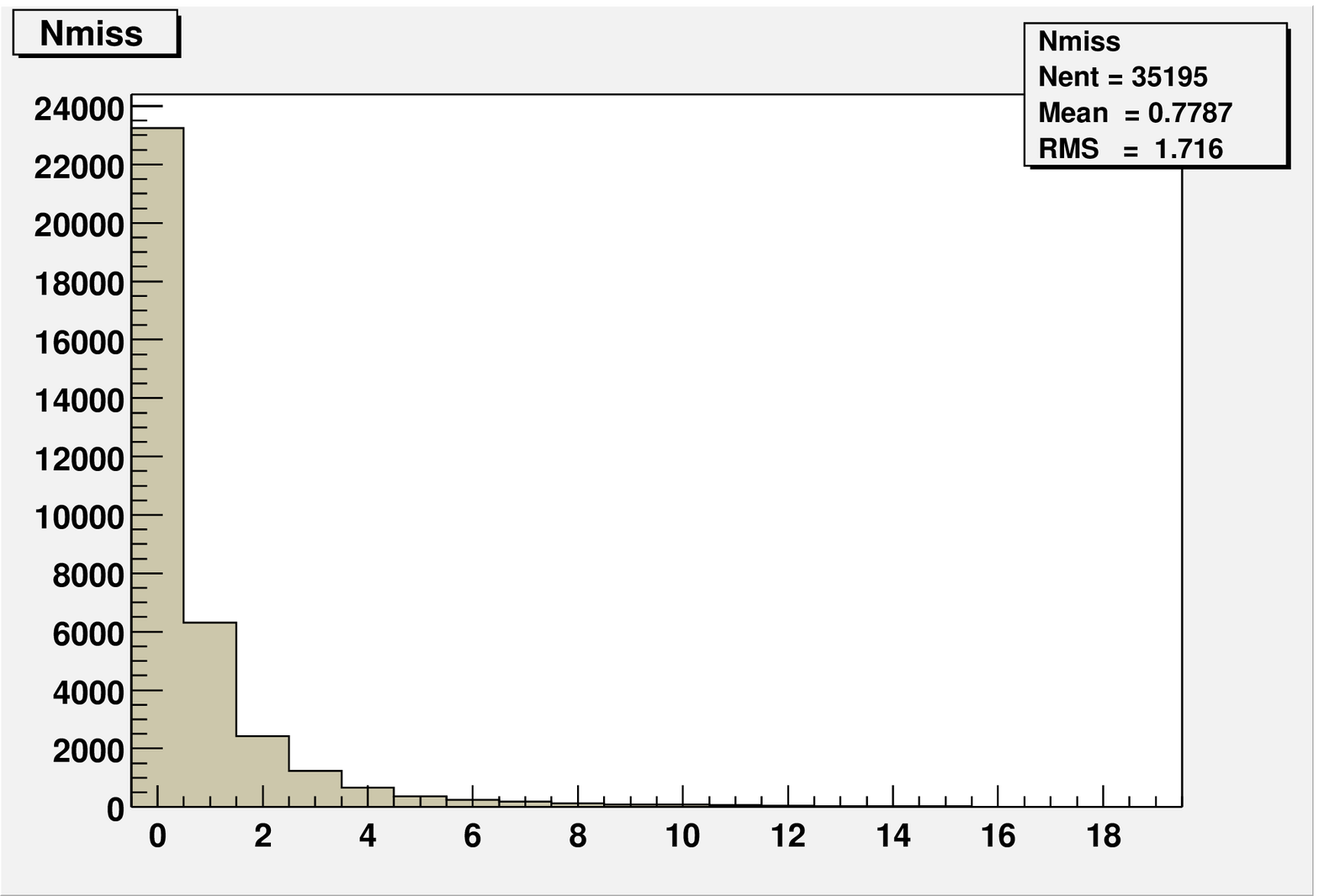,height=2.0in,clip=on}
                   \psfig{figure=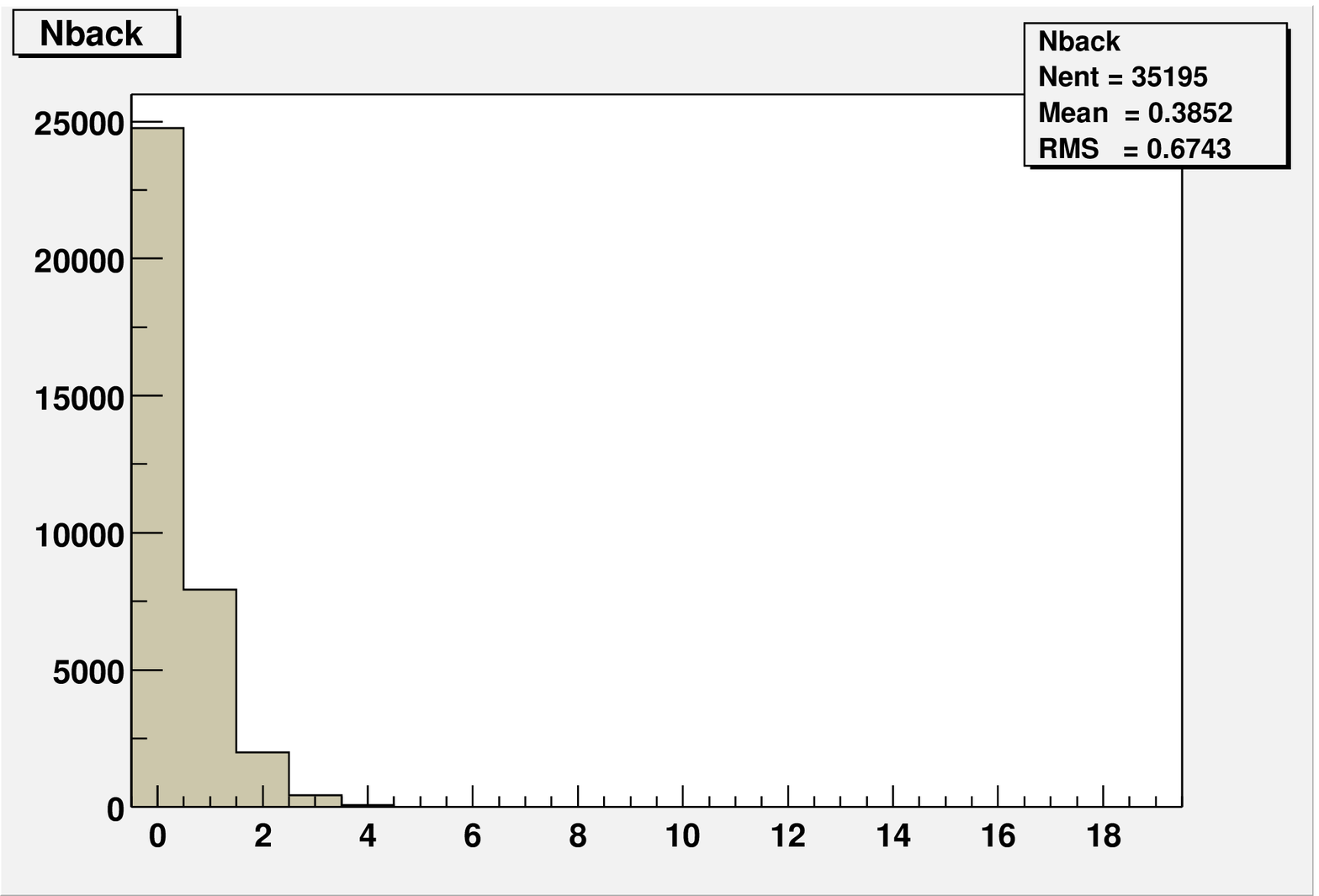,height=2.0in,clip=on}}}
  \caption{
  Distributions in the number of missing real hits and
  remaining background hits.
 }
\label{fig:Nback}
\end{figure}

Figure  ~\ref{fig:Nback}(a) represents the number of real
hits lost by the pattern recognition procedure. Some of the real
tracker hits (0.8 hits $\sim 2.7\%$) are lost due to the selection
procedure.

Figure  ~\ref{fig:Nback}(b) represents the number of background
hits remaining after the selection procedure. As one
can see from Figure ~\ref{fig:Nback}(b) the number of 
background hits remaining is 0.38 hits in comparison with the
primary 300 hits. So a total background suppression factor is
300/0.38 $\approx$ 800.

\section{Reconstruction based on Kalman filter }

Background studies presented in this subsection are based on the
application of the Kalman filter to hits selected by previous pattern 
recognition procedure.

At this stage the reconstruction is based on the hits selected by the
pattern recognition procedure described above.

In principle we could use the results obtained by DAF for the track
reconstruction but our analysis shows that an application of
a combinatorial Kalman filter to hits selected by DAF provides better
precision. 

In the last decade the Kalman filter (KF) approach \cite{kalman}
has been extensively exploited for track fitting in high energy
physics. This approach possesses the following
features for effective track fitting: \\
\noindent
$\bullet$ multiple scattering and energy losses are included in a natural way;\\
$\bullet$ a 3D trajectory is restored that approximates closely the real one; \\
$\bullet$ complex tracker geometries are handled in a simple way;\\
$\bullet$ N$\times $N matrix inversion, where N is the total number of measurements,
is avoided; \\
$\bullet$ control for error propagation is provided; \\
$\bullet$ trajectory is reconstructed progressively from one measurement to the next,
improving the precision  with each step; \\
$\bullet$ initial and final momenta of a particle crossing the
tracker are reconstructed. \\

The KF is very useful because it simultaneously finds and fits 
the track;
it is much more economical than the
conventional least-squares global fit. The KF is a ``progressive"
step by step method whose predictions are rather poor at the
beginning of the track at the first stage of filtering. Since a
state vector ${\bf{x}_k}=(x,y,t_x,t_y,1/p_{L})$ at point k (see
definitions in Appendix A) has five parameters we need
approximately $\simeq$ 6-7 straw hits to get good KF prediction
precision. The prediction step, in which an estimate is made for
the next measurement from the current knowledge of the state
vector, is very useful to discard noise signal and hits from other
tracks. Assuming the validity of the helix track model for each
step, the KF propagates the track in 3D space, from one 2D surface
to the next.

Below we will use the standard notations:

$\bf{x}_{k+1}^{k}$ is a prediction, i.e. the estimation of the
``future" state vector at position ``k+1" using all the ``past"
measurements up to and including ``k".

$\bf{x}_{k}^{k}$ is a filtered state vector, i.e. the
estimation of the state vector at position ``k" based upon all
``past" and ``present" measurements up to and including ``k".

The same notations will be held for the covariance matrix
$\bf{C}$, noise matrix $\bf{Q}$ and so on.

The Kalman filter algorithm can be divided into three major steps. \\

INITIALIZATION

The KF can start from arbitrary parameters and an infinite
covariance matrix but for track finding applications it is
significantly better to fix somehow the initial state if possible.

The initialization of forward and backward KF algorithms is quite
simple. It starts from the artificial point and initial
parameters calculated from the previous stage of reconstruction
procedure. The initial covariance matrix is empirically found to be diagonal
with matrix elements being much greater than the corresponding
uncertainties: $\bf{C}_{0}^{0}$ = (0.3,0.3,0.03,0.03,0.000003).\\

FILTERING

Once the KF is initialized it makes standard consequent steps. The
current state at the k-th step is defined by the state vector
$\bf{x}_k$, the state covariance matrix $\bf{C}_{k}^{k}$
and the current straw hit.
To take the (k+1)-th step it is necessary to:\\

\noindent
$\bullet$ update $\bf{x}_{k}^{k}, \bf{C}_{k}^{k}$ to take into account the ionization losses \\
$\bullet$ define the next hit object (up or down point) \\
$\bullet$ propagate the parameters and the covariance matrix
to: $\bf{x}_{k}^{k} \rightarrow \bf{x}_{k+1}^k$, $\bf{C}_{k}^{k} \rightarrow \bf{C}_{k+1}^k$\\
$\bullet$ update $\bf{C}_{k+1}^k$ to take into account
multiple scattering $\bf{C}_{k+1}^k \rightarrow \bf{C}_{k+1}^k + \bf{Q}_k$\\
$\bullet$ calculate the Kalman matrix $\bf{K}_{k+1}$ \\
$\bullet$ update the covariance matrix $\bf{C}_{k+1}^{k+1}$ \\
$\bullet$ calculate residuals $\bf{r}_{k+1}^{k+1} $ and their covariance matrices $\bf{R}_{k+1}^{k+1} $ \\
$\bullet$ calculate the incremental $\chi^2 =
(\bf{r}_{k+1}^{k+1})^T (\bf{R}_{k+1}^{k+1})^{-1} \bf{r}_{k+1}^{k+1}$ \\
$\bullet$ store all information defining the new state\\

SMOOTHING

In the standard Kalman filter algorithm the smoothing is a well
defined procedure. Smoothing allows one to obtain the best
estimate of the track parameters at any trajectory point using all
hit information accumulated during the KF propagation.

\begin{figure}[htb!]
\centerline{\hbox{%
\psfig{figure=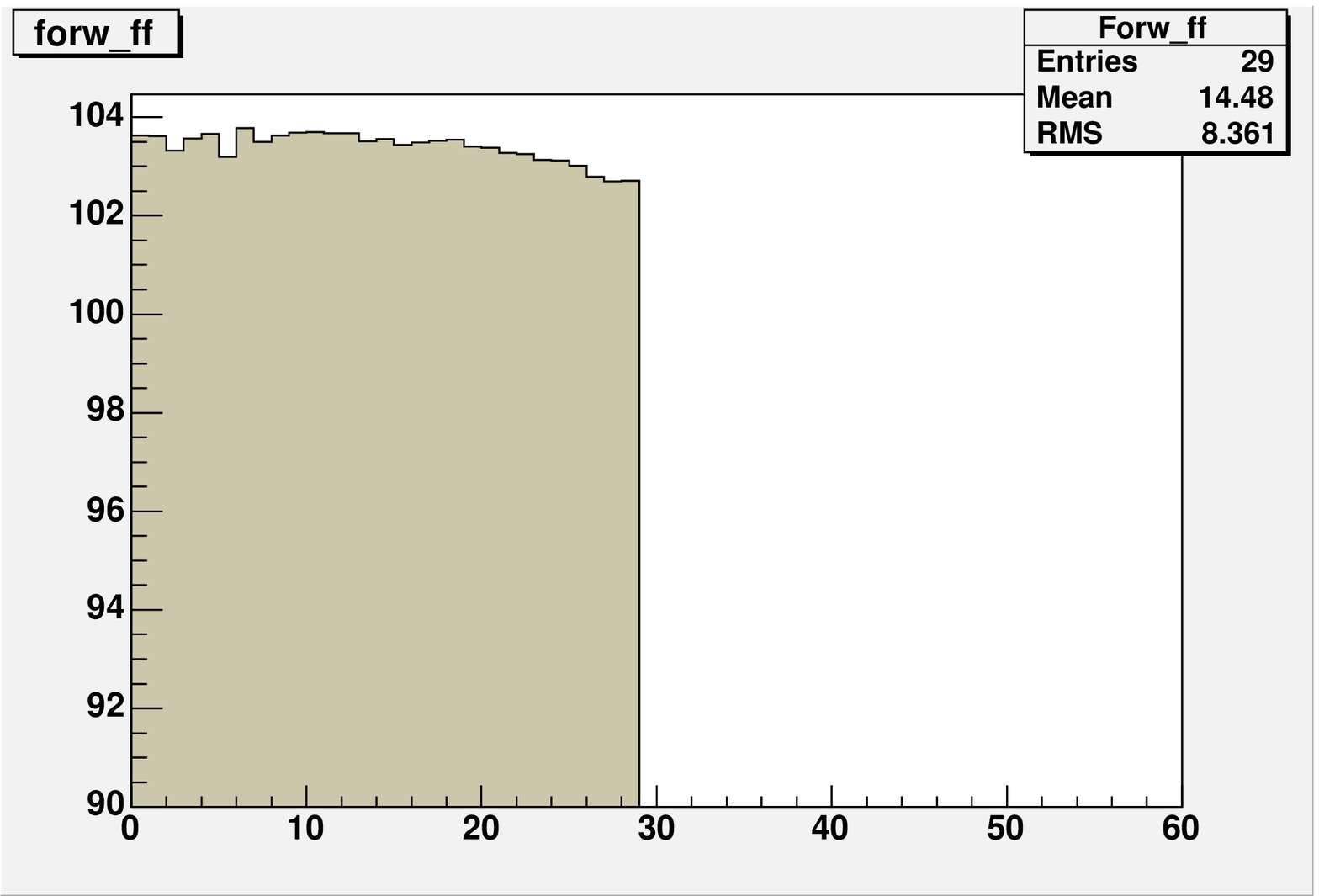,height=2.0in,clip=on}
 \psfig{figure=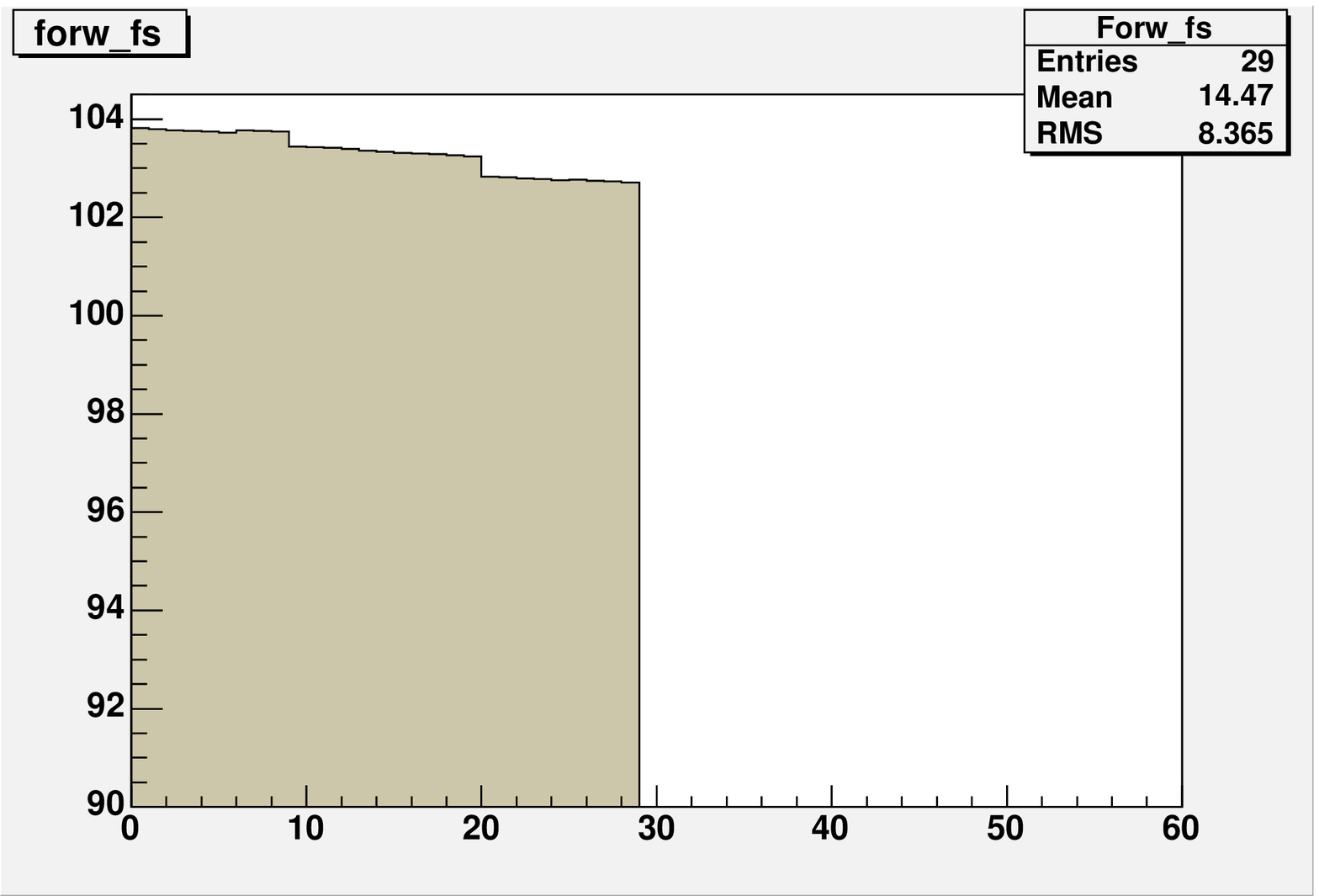,height=2.0in,clip=on}}}
\caption{
 The total momentum reconstructed by the
 forward Kalman filtering a) and by the smoothing b)
for each hit of a selected event. The y-axis is Ptot, the
x-axis is the z-ordered hit number.
 }
\label{fig:forw}
\end{figure}

Figure ~\ref{fig:forw} (a) displays the results of the KF
forward filtering for the total momentum Ptot
reconstruction at each tracker hit position. Figure
~\ref{fig:forw} (b) displays how the KF smoother, based on all
hit information accumulated during the KF filtering, improves the
total momentum Ptot reconstruction at each tracker hit
position.

\begin{figure}[htb!]
\centerline{\hbox{%
\psfig{figure=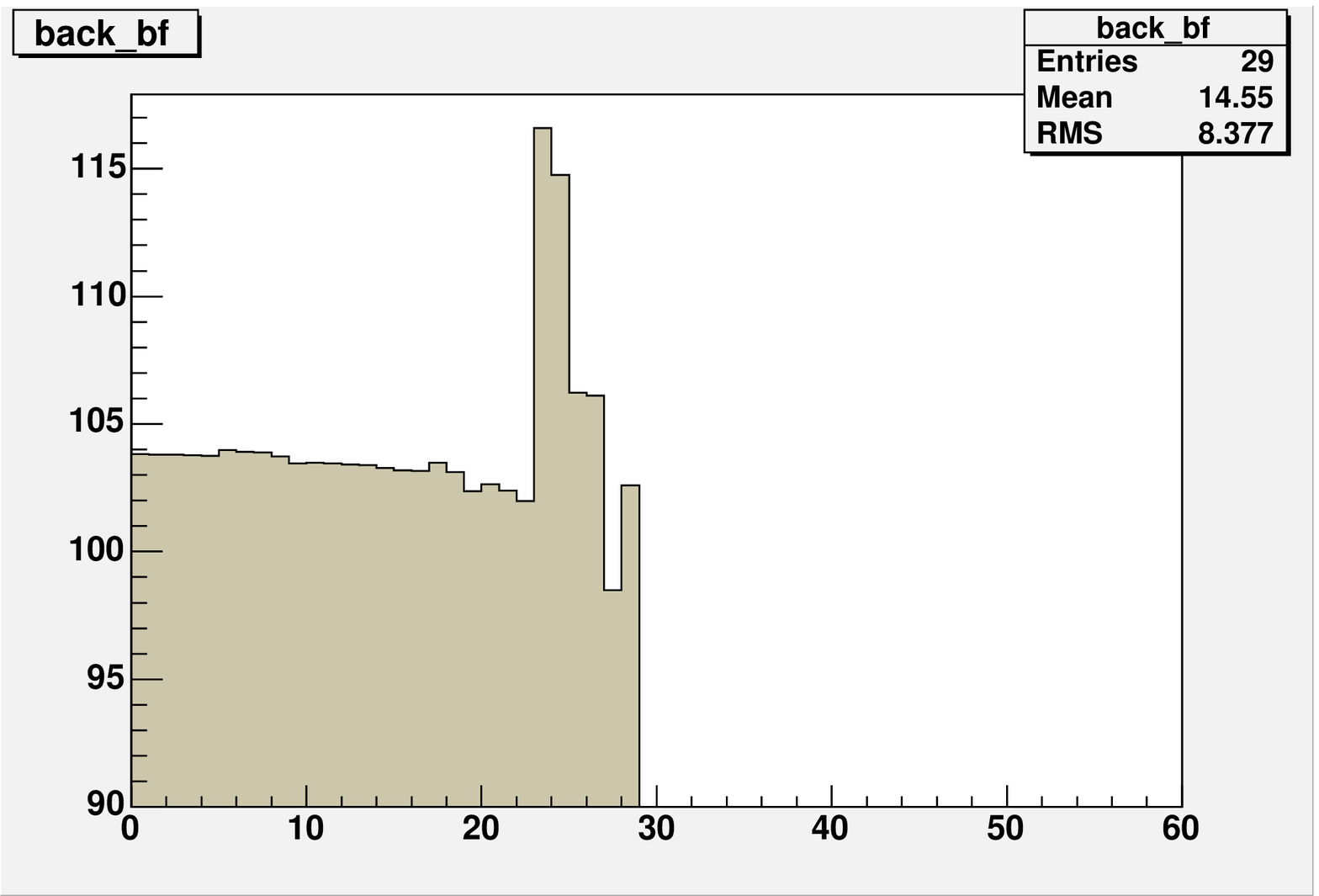,height=2.0in,clip=on}
 \psfig{figure=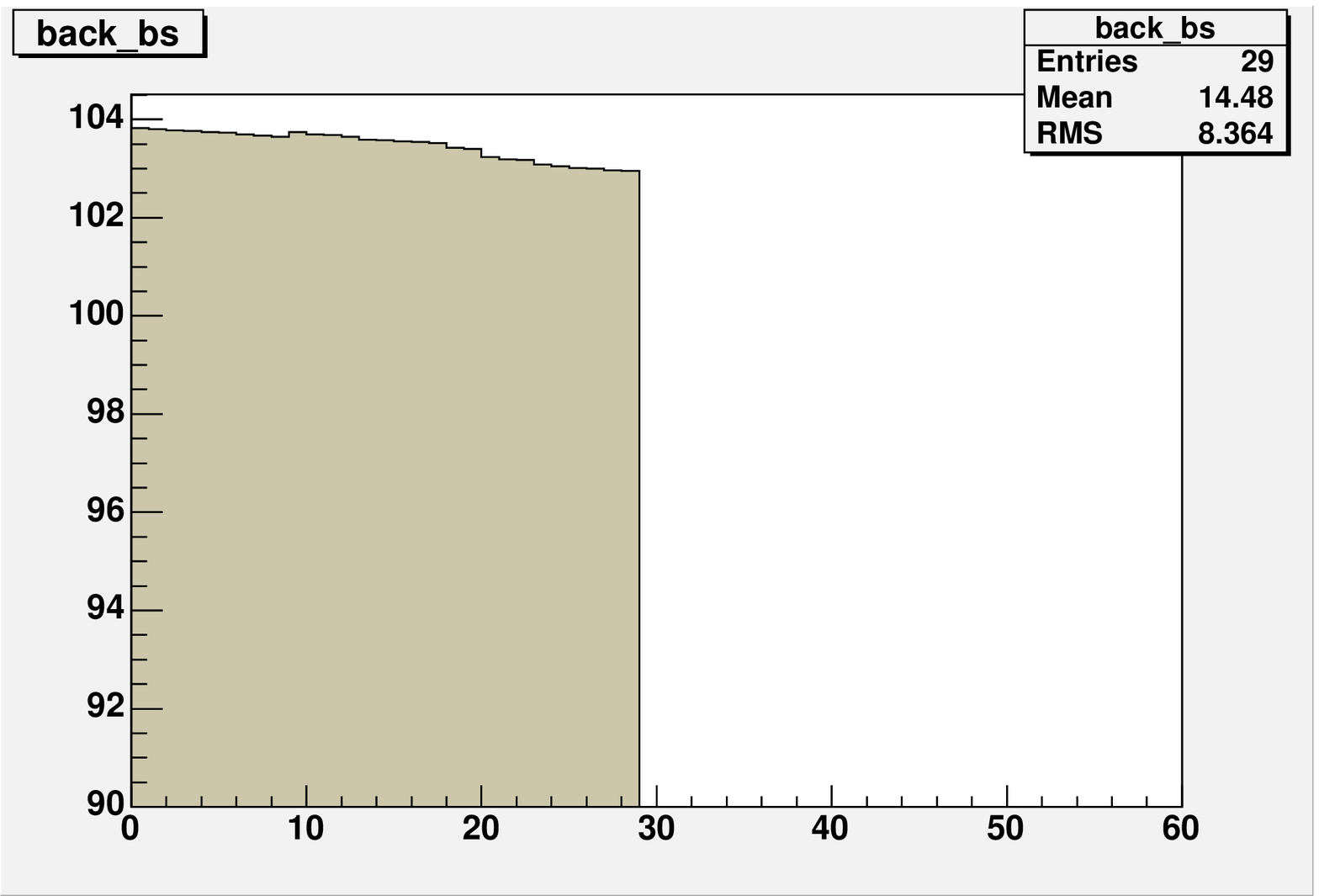,height=2.0in,clip=on}}}
\caption{ The total momentum reconstructed by the
 backward Kalman filtering a) and by the smoothing b)
for each hit of a selected event. The y-axis is Ptot, the
x-axis is the z-ordered hit number.
  }
\label{fig:back}
\end{figure}

Figure ~\ref{fig:back} (a) displays the results of the KF
backward filtering for the total momentum Ptot 
reconstruction at each tracker hit position. Figure
~\ref{fig:back} (b) displays how the KF smoother, based on all
hit information accumulated during the KF filtering, improves the
total momentum Ptot reconstruction at each tracker hit
position.

The Kalman filter approach is effective in the resolving 
of up - down ambiguity. By
applying the Kalman filter at each step up and down points are
considered and the point providing the best $\chi^{2}$ for the
trajectory is selected as the true point. This procedure is
approximately linear in the number of tracker hits in comparison with
the a combinatorial search, which is not feasible.

At this reconstruction stage due to left-right ambiguity for the
straw drift chamber we have a set of true and mirror  points for
each straw hit. As discussed above the reconstruction procedure
for each tracker hit defines true and mirror  points and one of
them is close to the real point with high precision $\simeq$
25 $\mu$m. 
So we can formulate our goal as to find a true point
combination for N straw hits in the presence of N mirror points.

The procedure to find the best approximation to the true point
combination for N straw hits in the presence of N mirror  points
for one track from a muon-electron conversion event
is based on the following steps:\\

$\bullet$ chose the first eight straw hits and built $2^8$ (256)
possible hit combinations corresponding to up and down points.
The KF forward and backward procedures described above are applied
to these combinations. Only those combinations which satisfy a
rather loose $\chi^2$ cut, $\chi^2 <$ 30,
are retained (typically about 10 out of the initial 256 combinations);\\

$\bullet$ make a loop for all retained combinations with fixed up
and down points for the first eight straw hits. For the 9th and
higher straw hit, the up and down point choices are take into
account and the point  with minimal incremental $\chi^2$ for
this point is selected for the further KF propagation step. If
both incremental $\chi^2$ satisfy the cut $\chi^2 < $ 10 the
second point is stored in the stack to make an iterative loop.
A single combination of all possible steps defines a candidate
track. At this stage on average 45 candidate tracks are stored
in the stack per event; \\

$\bullet$ make a loop for all combinations from the stack. For
each new hit added to the hits restored from the stack again the
two up and down point choices are taken into account and the
point with the minimal incremental $\chi^2$ for this point
is selected for the further KF propagation step; \\

$\bullet$ select up and down track combinations with the minimal $\chi^2$ for the track; \\

$\bullet$ select a track satisfying the cut $\chi^2 < $ 70; \\

$\bullet$ select a track with the difference between the forward
(Pin\_f) and backward  (Pin\_b)
 reconstructed input momentum satisfying the cut $\vert Pin\_f - Pin\_b \vert <$ 0.7 MeV/c.  \\

   The Kalman filter reconstructs a trajectory of a particle
in three dimensions. The trajectory is bent
each time it crosses a tracker plane due to multiple 
scattering. Therefore, the reconstructed track is a set of
helices that intersect at the  planes.
This is the track followed by the particle. 
   
\begin{figure}[htb!]
\centerline{\hbox{\psfig{figure=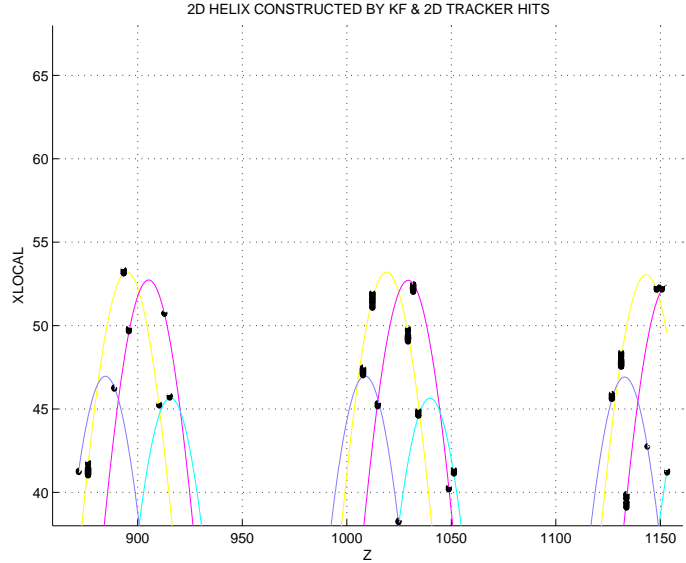,height=3.0in,clip=on}}}
 \caption{
 Plot of real + background tracker hits after the DAF selection procedure.
 }
\label{fig:ev2}
\end{figure}

   Figure ~\ref{fig:ev2}  displays a 2D projection of 3D trajectory
reconstructed by the Kalman filter for the sample event.
As above in this figure all 12 views were joined in one 
for all 18 tracker modules. The 2D trajectory is shown 
only for sensitive area of the tracker and for each view 
the trajectory is in a different color. 
For the sample event real hits are in four 
tracker's views and the reconstructed
lobes for these views are clearly seen in the figure.

   Due to a scale in this figure the 2D trajectory looks
as an ideal sine curve and tracker hits look like spots of
different size. In order to see a detailed behavior of the
trajectory and hit positions a dynamical zoom is applied 
to a rectangular region indicated in  Figure ~\ref{fig:ev2} 
in the x-range 49-50 cm and
z-range 1029-1029.5 cm. 
   
\begin{figure}[htb!]
\centerline{\hbox{\psfig{figure=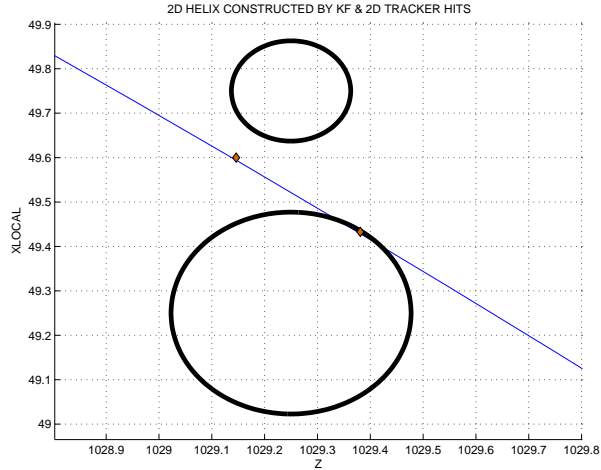,height=2.5in,clip=on}}}
 \caption{
 Zoom enlargement of the region in Figure ~\ref{fig:ev2}.
 }
\label{fig:ev21}
\end{figure}

Figure ~\ref{fig:ev21} demonstrates the magnified region of the tracker.
Blue line in the figure represents the reconstructed trajectory.
Two circles represent two hits in  chamber. 
These circles look like ellipses due to different axis scales.

Radii of the circles are directly
proportional to the drift times. 
For illustrative 
purposes two nearest to the straw center points corresponding to the real
trajectory   obtained in Monte Carlo
simulation are shown in the Figure in the form of diamonds
~ (note that in the pattern recognition procedure only 
radii were used but not these points).
The measurement uncertainty was taken into account assuming
that circle radii are distributed normally about the simulated 
radii with $\sigma = 200 \mu$m. For this reason the position 
of one of the  nearest points is not on a circle in Figure ~\ref{fig:ev21}.
  
In Figure ~\ref{fig:ev21} due to the corresponding scale the trajectory 
looks like a straight line. It is tangent to one of two circles
obtained on the basis of drift time.  
In the region under consideration the deviation of the 
trajectory from the  nearest point is less than 0.2 mm. 
The change in the direction of the trajectory due to 
multiple scattering can not be seen in the Figure because of 
the smallness of the average angle of the scattering.        

\begin{figure}[htb!]
\centerline{\hbox{\psfig{figure=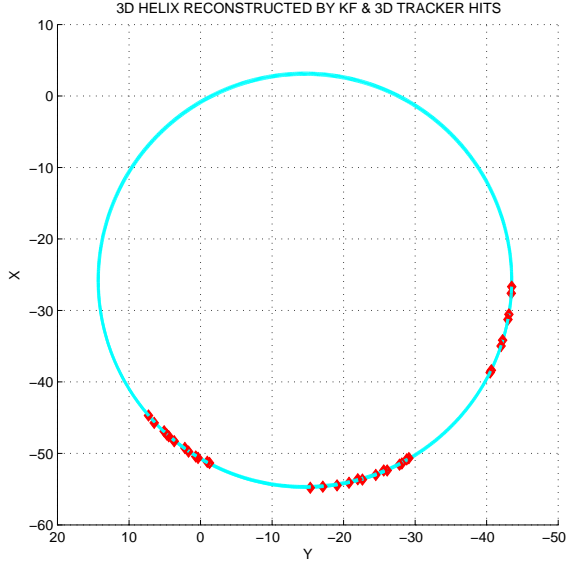,height=3.0in,clip=on}}}
 \caption{
 The transverse projection of the 3D trajectory reconstructed for the sample event.
 }
\label{fig:fig_3D3}
\end{figure}

Figure ~\ref{fig:fig_3D3}
shows transverse xy-projection of the 
trajectory for the sample event. 
In this projection the trajectory looks approximately 
as a circle.

\begin{figure}[htb!]
\centerline{\hbox{\psfig{figure=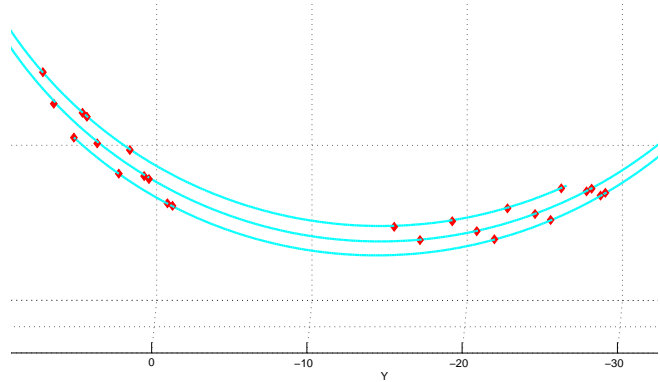,height=2.0in,clip=on}}}
 \caption{
 Zoom enlargement of the region in Figure ~\ref{fig:fig_3D3}.
 }
\label{fig:fig_3D4}
\end{figure}

However if a specific region of the tracker is magnified 
by the dynamical zoom one can see in Figure ~\ref{fig:fig_3D4} 
that the shape of the
circle is distorted due to multiple scattering and energy
loss. More than two turns of trajectory are clearly 
seen in the Figure.

 Note that by using the KF filter a momentum of a particle can be
 reconstructed at any point of the tracker. For our purposes the most
 important is the momentum of a particle entering the tracker,
which in the following we will call  the input momentum.

\begin{figure}[htb!]
\centerline{\hbox{\psfig{figure=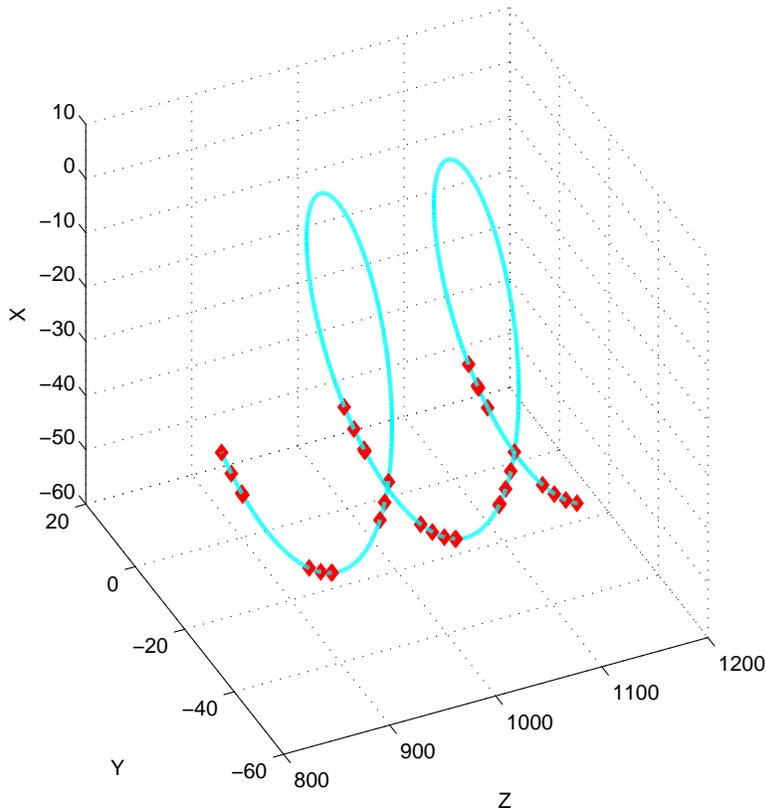,height=4.5in,clip=on}}}
 \caption{
 3D trajectory reconstructed for the sample event.
 }
\label{fig:fig_3D1}
\end{figure}

Figure ~\ref{fig:fig_3D1}
demonstrate 3D trajectory  reconstructed for the 
sample event. 
The trajectory looks in this scale as a helix, but we remind that it 
consists of many helix parts.
Also in the Figure tracker's hits 
generated by Monte Carlo simulation program are shown.

\begin{figure}[htb!]
\centerline{\hbox{%
\psfig{figure=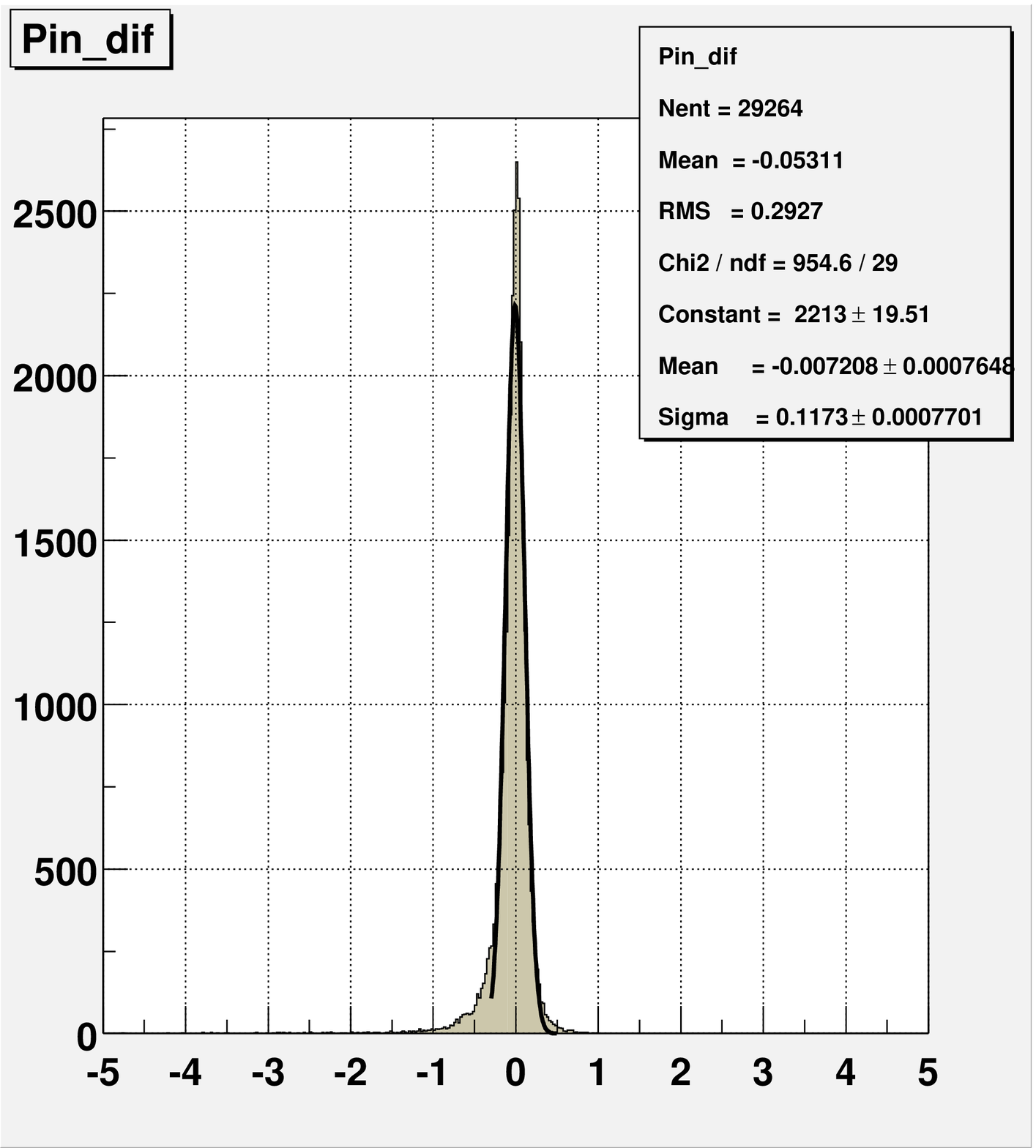,height=3.0in,clip=on}
 \psfig{figure=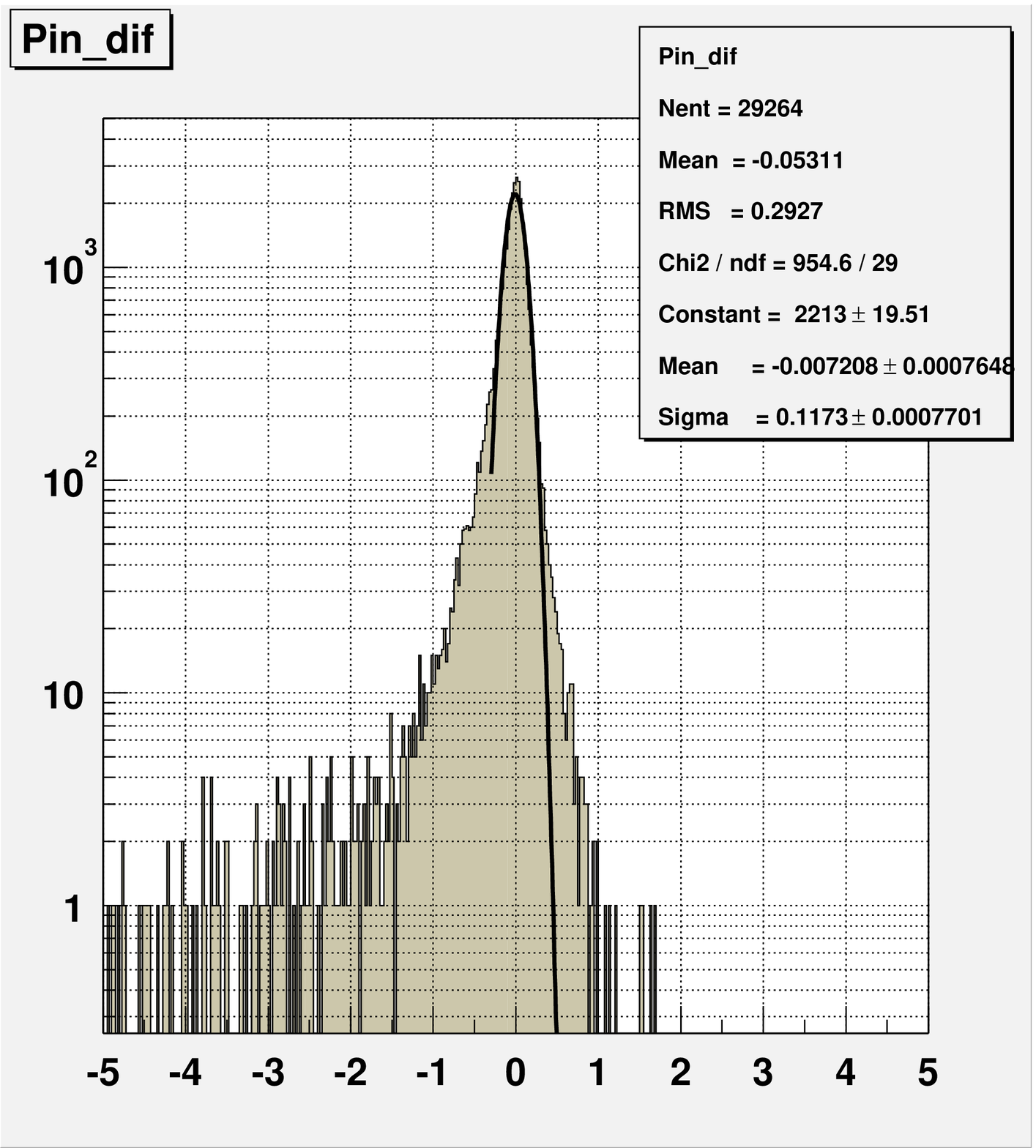,height=3.0in,clip=on}}}
\caption{
 Distribution in the difference between the input reconstructed momentum based on
 the Kalman filter and the simulated input momentum with background.
 }
\label{fig:pin_difb}
\end{figure}

 The distribution in the difference between the initial momentum (Pin\_f) 
reconstructed by
 the Kalman filter and the generated initial momentum (Pin) is shown in
Figure ~\ref{fig:pin_difb} in linear (a) and logarithmic (b) scale.
According to this distribution the intrinsic tracker resolution is
$\sigma$ = 0.12 MeV/c if one fits the distribution by a Gaussian in the range
-0.3 - 0.7 MeV/c.

\begin{figure}[htb!]
\centerline{\hbox{%
\psfig{figure=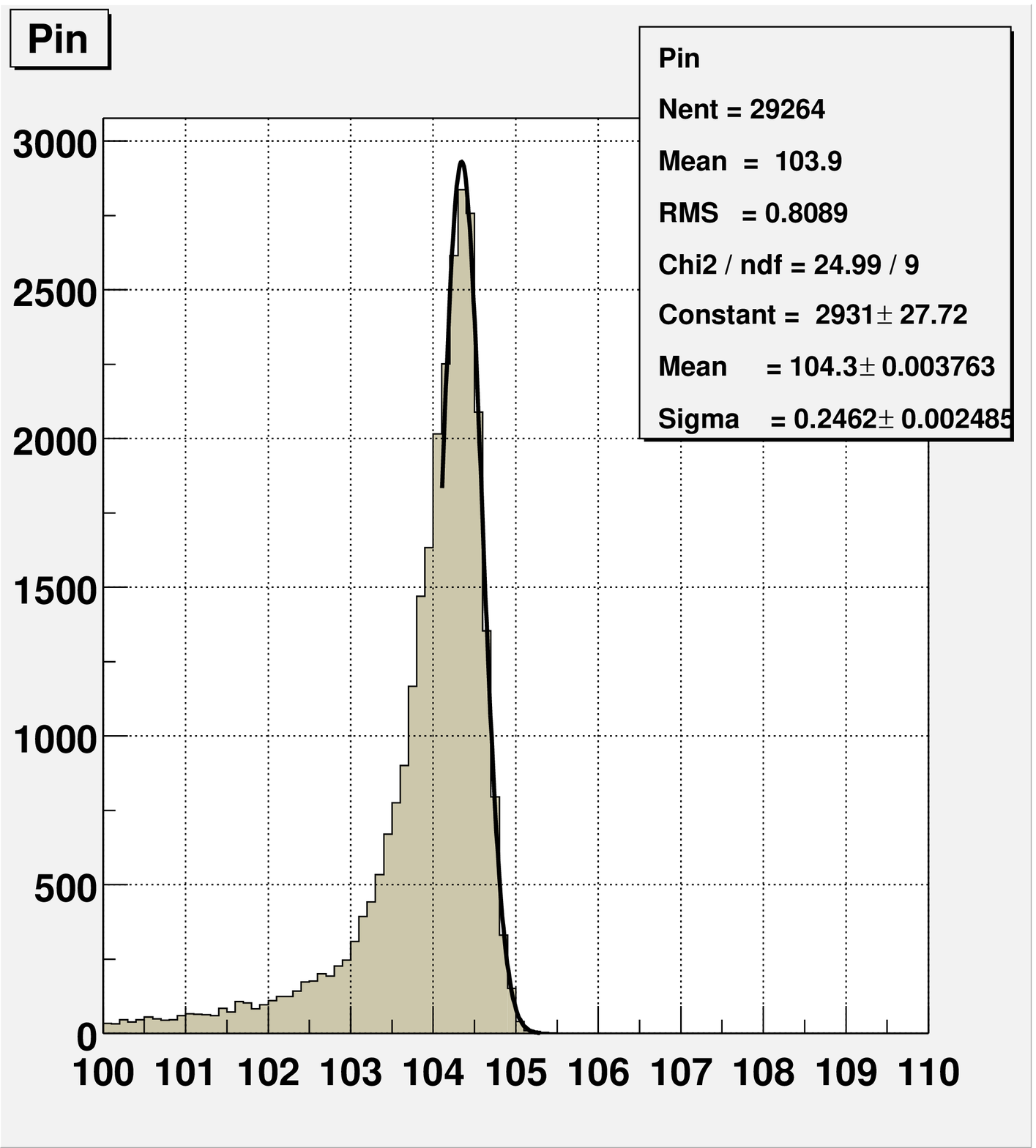,height=3.0in,clip=on}
 \psfig{figure=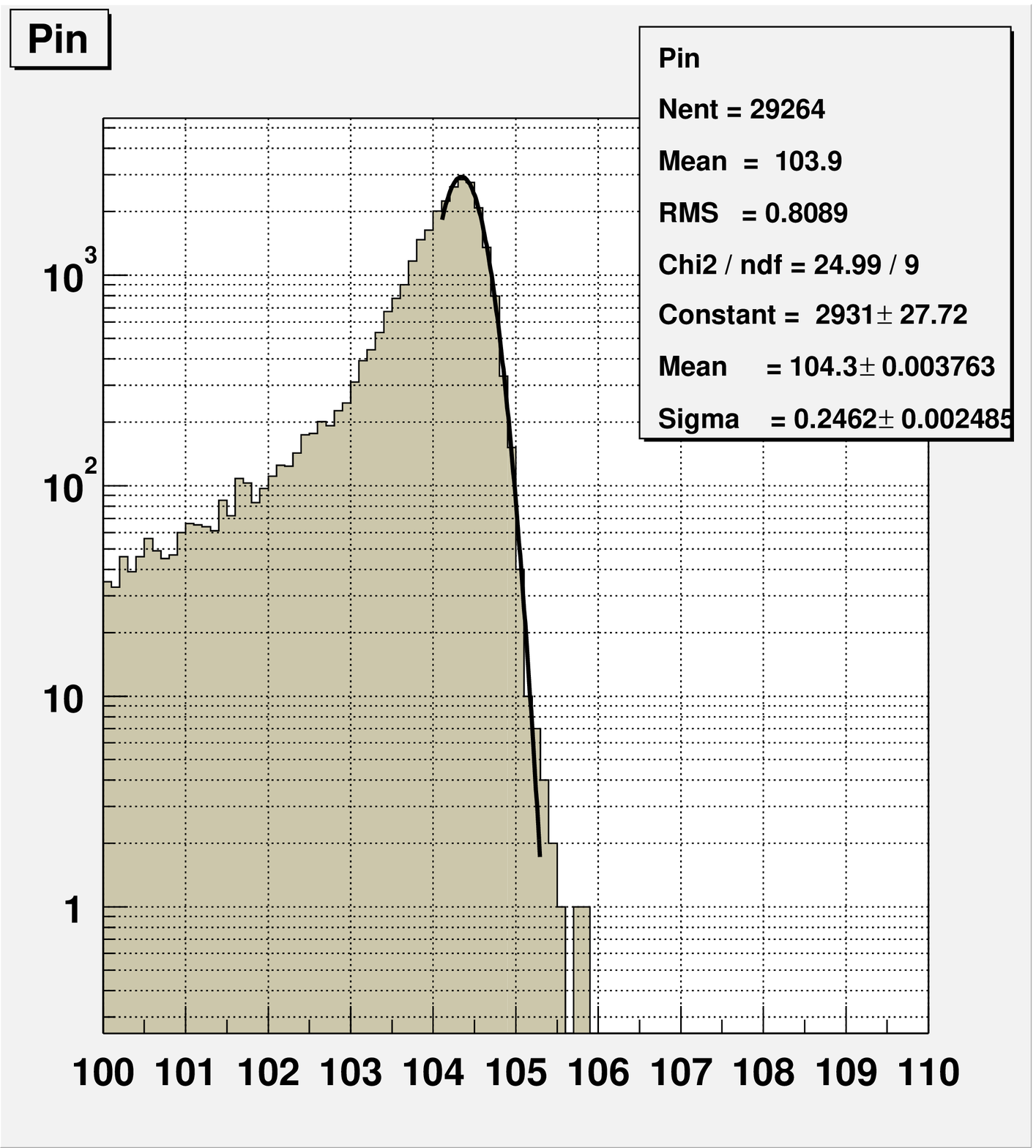,height=3.0in,clip=on}}}
\caption{
 Distribution in the input momentum reconstructed by the
 Kalman filter  with background.
 }
\label{fig:pin_kalb}
\end{figure}

Figure ~\ref{fig:pin_kalb} shows a distribution in the input
momentum (Pin\_f) reconstructed by the Kalman filter 
in linear (a) and logarithmic (b) scale with background.  
This distribution is characterized by the standard deviation $\sigma$ = 0.25 MeV/c of the
reconstructed input momentum  for a Gaussian fit
in the range 104-106 MeV/c.

Note that the trajectory
reconstructed by the Kalman filter consists of
many helix parts.  
The reconstructed input momentum resolution by the  
Kalman filter is $\sigma$ = 0.25 MeV/c. This resolution significantly
better than the resolution $\sigma = 0.35$ MeV/c obtained 
by a single helix fit.

The overall reconstruction acceptance is 22.1 $\%$ for muon conversion
events with the momentum above a threshold momentum of 103.6
MeV/c.

Comparing these results with the results of the reconstruction
without background we get the difference in tracker
resolution 1.5 \% and the difference 2.7 \% in overall acceptance
(see Appendix C).
Therefore the tracker resolution and overall acceptance are not
affected significantly at the considered background level 
(explained in Section 3).

A summary of the critical selection criteria used in the momentum
reconstruction is shown in Table ~\ref{table:tab5}. The
efficiencies are for the selection criteria applied in consecutive
order.

An overall acceptance for muon conversion events with momentum
above threshold momentum $P_{th}$ is 22.1\% .
We define a threshold
momentum,
above which events are considered as the useful ones by
 $P_{th} = P_{max} - \Delta $, where $P_{max}$ = 104.3 MeV/c
the  most probable reconstructed momentum. If $\Delta $ = 0.7 MeV/c
is chosen then $P_{th}$ = 103.6 MeV/c.

\begin{center}
\begin{table}[htb!]
\caption {A summary of the reconstruction selection criteria }
\begin{tabular}{|l|c|}
\hline
 Selection criterion & Efficiency\\
\hline
Calorimeter energy above 80 MeV & 0.53\\
Required pitch angle at the tracker & 0.86\\
At least 15 hits in the tracker & 0.87\\
Position match in the calorimeter & 0.95\\
Requirements on fit quality & 0.79\\
Detected energy above 103.6 MeV & 0.74\\
\hline
Overall acceptance & 0.22\\
\hline
\end{tabular}
\label{table:tab5}
\end{table}
\end{center}

The main factors entering into the experimental sensitivity are
the running time, the proton intensity, the probability per proton
that a $\mu$ is produced, transported and stopped in the stopping
target, the fraction of stopped muons that are captured (as opposed to decay),
the trigger efficiency and the tracker reconstruction acceptance.
We do not include in this table loss of events due to accidental
cosmic ray vetoes, dead-time losses and losses due to straw
chamber inefficiencies, all of which are expected to be small.
According to our analysis taking into account the straw efficiency
97\% the overall acceptance is reduced from 22.1\% to 21.7\% .

Table ~\ref{table:tab6} shows expected MECO sensitivity for a one
year ($10^7$ s) run.

\begin{center}
\begin{table}[htb!]
\caption {A summary of the expected MECO sensitivity.}
\begin{tabular}{|l|r|}
\hline
Running time (s)& $10^7$\\
Proton flux (Hz)& $4 \cdot 10^{13}$\\
\hline
Probability of $\mu$/p transported and stopped in target & 0.0025\\
$\mu$ capture probability & 0.6 \\
Fraction of $\mu$ which are captured in time window & 0.49 \\
Trigger efficiency and the selection criteria & 0.22 \\
\hline
Detected events for $R_{\mu e} = 10^{-16}$ & 6.5\\
\hline
\end{tabular}
\label{table:tab6}
\end{table}
\end{center}

Muon DIO events are the most important background for the experiment.
The main background from muon DIO events in the presence of background tracker hits
was simulated and reconstructed.
Based on the simulated DIO events
in the momentum range
above 100 MeV/c the track pattern recognition and momentum reconstruction
were performed in the presence of background tracker hits
(protons, neutrons, photons, DIO) by applying the  selection criteria
discussed above.

\begin{figure}[htb!]
\centerline{\hbox{\psfig{figure=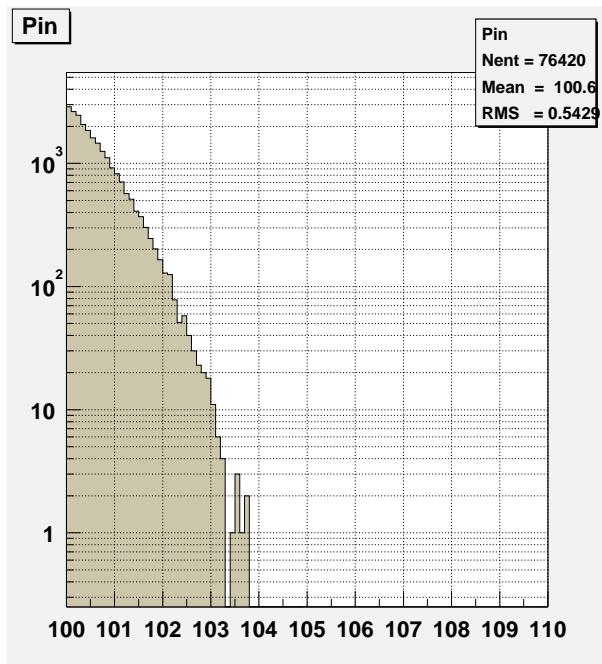,height=3.5in,clip=on}}}
  \caption{
  Distributions of DIO reconstructed momentum in the presence of the background 
 above 100 MeV/c .
 }
\label{fig:dio}
\end{figure}

The number of primary DIO events  simulated in the momentum range above 100 MeV/c
was  10 times more
than expected and 3 background events were found (see Figure ~\ref{fig:dio}).
So the background is expected to be 0.3 events.

It is important to note that input reconstructed momenta for 
these three background events are very close to  
simulated input momenta. The input simulated 
momenta 103.48, 103.6 and 103.74 MeV/c for these events have to be
compared with input reconstructed momenta 103.6, 103.76 and 103.8 MeV/c, 
respectively.

At the present level of pattern recognition and momentum
reconstruction studies we can expect the background from DIO
events in the range above 100 MeV/c $\sim $0.3 events compared 
to 6.5 signal events  for $R_{\mu e} = 10^{-16}$.

\section{Conclusion}

A study of the impact of background on the performance of the 
transverse tracker proposed for the MECO experiment is presented.
Background from capture protons, neutrons and photons, and from
muon decay in orbit was generated using GEANT3. 
The effective average straw tube rate from these sources was 800 kHz
at the proposed muon beam intensity of $2\times 10^{11} \mu^{-}/sec$. 
   
A pattern recognition
procedure based on a Kalman filter technique was developed to 
suppress background and assign hits to tracks.
In the first stage of this procedure, straw hit center 
coordinates, without drift time information, were used to reduce 
the background by a factor $\sim$ 130. In the second stage, the full 
drift time information
and a deterministic annealing filter were used to obtain an 
additional six-fold suppression. The total suppression of 800 reduces
the number of background hits on average from an initial value of
300 to approximately 0.4 per event. About 0.8 hits of the 29 real hits
typically recorded, or 2.7 $\%$, are lost in the process.

It was found that in the presence
of background the
resolution of the tracker is $\sigma = 0.12$ MeV/c and the overall 
setup acceptance for muon conversion
events with momentum above the threshold momentum 103.6 MeV/c is 
about 22 $\%$. 
At the considered background level the tracker resolution and the 
overall acceptance are not affected significantly by presence of 
the background: the tracker resolution is changed by 1.5 \% and the 
overall acceptance by 2.7 \%.

Additional constraints on the background, not considered here, may be
imposed if the drift time measurement is supplemented by a 
measurement of the pulse amplitude at the anode wire. A straightforward, 
crude measurement of the amplitude is sufficient to reduce significantly
the background from heavily ionizing particles, i.e., the capture
protons that comprise 30 \% of the background hits in the above study. 
We estimate too that a significant background suppression, of 10-20,
would be achieved if resistive anode wires were used in place of 
conducting wires to obtain a measurement of the hit position along
the wire. This would improve the resolution as well. 

At the present level of pattern recognition and momentum
reconstruction studies the background from DIO events above 100 MeV/c
is about 0.3 events. This is to be compared 
to 6.5 signal events  for $R_{\mu e} = 10^{-16}$.

The study carried out shows that the developed procedures of pattern 
recognition and momentum reconstruction in the case of the transverse tracker 
provide a required precision for lepton number violation search at a sensitivity
level about $10^{-17}$.

We wish to thank A. Mincer, P.Nemethy, J.Sculli and one of us (R.K.) thanks
W.Willis for fruitful discussions and helpful remarks.

\newpage

\section*{APPENDIX\ A. KALMAN FILTER AND ITS\ APPLICATION\ TO\ TRACK\ FITTING}

The Kalman filter is an algorithm that processes measurements to deduce an
optimum estimate of the past, present, or future state of \ a dynamic system
by using a time sequence of measurements of the system behavior, plus a
statistical model that characterizes the system and measurements errors,
plus initial condition information.

The Kalman filter addresses the general problem of trying to
estimate at different points ($1\leq k\leq n$) the state
$\bf{x}_{k}$ of a discrete process that is governed by the
linear stochastic difference equation

\begin{equation}
\bf{x}_{k}=\bf{F}_{k-1}\bf{x}_{k-1}+\bf{w}_{k-1}
\label{1ap}
\end{equation}

with a measurement $\bf{m}_{k}$ that is

\begin{equation}
\bf{m}_{k}=\bf{H}\bf{x}_{k}+\bf{\varepsilon} _{k}.
\label{2ap}
\end{equation}

The system equation (\ref{1ap}) is not deterministic since the
track experiences stochastic processes such as multiple
scattering, bremsstrahlung, etc. These processes are taken into
account by the process noise $\bf{w}_{k}$ .
$\bf{\varepsilon} _{k}$ represents the measurement noise.
$\bf{w}_{k}$ and $\bf{\varepsilon} _{k}$ are assumed to be
independent of each other with zero expectation values:

$E\{\bf{w}_{k}\}=0,\qquad
cov\{\bf{w}_{k}\}=\bf{Q}_{k}$ , \qquad $1\leq k\leq n,$

$E\{\bf{\varepsilon} _{k}\}=0,\qquad cov\{\bf{\varepsilon} _{k}\}=\bf{V}_{k}$ , \qquad $%
1\leq k\leq n,$

where $\bf{Q}_{k}$ and $\bf{V}_{k}$  are process noise and
measurement noise covariances, respectively.

Eq.(\ref{1ap}) in the absence of the last term is the standard
equation of motion with a propagator $\bf{F}_{k-1}$ (transport
matrix). Note that at the moment $\bf{F}_{k-1}$ is assumed to
be constant.

Regarding a track in space as a dynamic system the filtering
technique is applied to the track fitting. For example, in the
case of a particle moving in magnetic field this can be done
naturally by
identifying the state vector $\bf{x}_{k}$ of the dynamic system with a vector $%
\bf{x}_{k}=(x,y,\tan \theta _{x},\tan \theta _{y},1/p_{L})$ of
5 parameters uniquely describing the track at each point of the
trajectory. The $\bf{F}$ matrix propagates the state vector on
one plane to the state vector on the next plane combining position
information with directional information. The transport matrix
implicitly contains information about a gap between planes.

In general the set of parameters  $\bf{x}_{k}$ is not measured
directly; only a function of $\bf{x}_{k}$ ,
$\bf{H}\bf{x}_{k}$ is observed. For example, in the case
of the transverse tracker one does not measure $\bf{x}_{k}$
but $x\prime =x\cos \alpha +y\sin \alpha $ in the chamber
coordinate system which corresponds to

\begin{equation}
\bf{H}=(\cos\alpha ,\sin \alpha ,0,0,0)  \label{2ap1}
\end{equation}

There are three types of operations to be performed in the analysis of a
track.

\begin{itemize}
\item \textbf{Prediction} is the estimation of the ``future" state
vector at
position ``k" using all the ``past" measurements up to and including ``k-1". $%
\bf{x}_{k}^{k-1}$ is a prediction (a priori state estimation).

\item \textbf{Filtering }is the estimation of the state vector at
position ``k" based upon all  ``past" and ``present" measurements
up to and including ``k". $\bf{x}_{k}^{k}$ is a filtered state
vector (a posteriori state estimation).

\item \textbf{Smoothing }is the estimation of the ``past" state
vector at
position ``k" based on all ``n" measurements taken up to the present time. $%
\bf{x}_{k}^{n}$ is a smoothed state vector.
\end{itemize}

The first step to estimate $\bf{x}_{k}$ is \textbf{the
prediction} (time update):

\begin{equation}
\bf{x}_{k}^{k-1}=\bf{F}_{k-1}\bf{x}_{k-1}^{k-1}
\label{3ap}
\end{equation}

\begin{equation}
\bf{C}_{k}^{k-1}=\bf{F}_{k-1}\bf{C}_{k-1}^{k-1}\bf{F}_{k-1}^{T}+\bf{Q}_{k-1}
\label{4ap}
\end{equation}

where Eq.(\ref{3ap}) projects the state ahead and  Eq.(\ref{4ap}) projects
the error covariance ahead.

\textbf{The filtered estimate} (measurement update)
$\bf{x}_{k}^{k}$ is calculated as a weighted mean of the
prediction and the observation:

\begin{equation}
\bf{K}_{k}=\bf{C}_{k}^{k-1}\bf{H}_{k}^{T}[\bf{H}_{k}\bf{C}_{k}^
{k-1}\bf{H}_{k}^{T}+\bf{V}_{k}]^{-1}
\label{5ap}
\end{equation}

\begin{equation}
\bf{x}_{k}^{k}=\bf{x}_{k}^{k-1}+\bf{K}_{k}[\bf{m}_{k}-\bf{H}_{k}\bf{x}_{k}^{k-1}]
\label{6ap}
\end{equation}

\begin{equation}
\bf{C}_{k}^{k}=[\bf{I}-\bf{K}_{k}\bf{H}_{k}]\bf{C}_{k}^{k-1}.
\label{7ap}
\end{equation}

Eq.(\ref{5ap}) computes the Kalman gain matrix defining the
correction to the predicted state due to the current observation.
Eq.(\ref{6ap}) updates the prediction with the measurement and
Eq.(\ref{7ap}) updates the error covariance. The error covariance
may be also expressed in a computationally superior form

\begin{equation}
\bf{C}_{k}^{k}=[\bf{I}-\bf{K}_{k}\bf{H}_{k}]
\bf{C}_{k}^{k-1}[\bf{I}-\bf{K}_{k}\bf{H}_{k}]^{T}+\bf{K}_{k}\bf{V}_{k}\bf{K}_{k}^{T}.
\end{equation}

The filtering is a recursive operation. The prediction step and
the filtering step are repeated for the next plane proceeding
progressively from plane ``1" to plane ``n". The state vector at
the last filtered point contains always the full information from
all points.

At each step one can calculate the filtered residuals
$\bf{r}_{k}^{k}$ , the covariance matrix of the filtered
residuals $\bf{R}_{k}^{k}$\ and the filtered $\chi ^{2}$:

\[
\bf{r}_{k}^{k}=\bf{m}_{k}-\bf{H}_{k}\bf{x}_{k}^{k}
\]

\[
\bf{R}_{k}^{k}=\bf{V}_{k}-\bf{H}_{k}\bf{C}_{k}^{k}\bf{H}_{k}^{T}
\]

\[
\chi
_{k}^{2}=\bf{r}_{k}^{kT}(\bf{R}_{k}^{k})^{-1}\bf{r}_{k}^{k}
\]

where $\chi _{k}^{2}$ is $\chi ^{2}$ - distributed with
dim($\bf{m}_{k}$) degrees
of freedom. The total $\chi ^{2}$ of the track is given by the sum of the $%
\chi _{k}^{2}$ contributions for each plane.

The system of equations defining the Kalman filter represents an
asymptotically stable system, and therefore, the estimate of the state
vector $\bf{x}_{k}^{k}$ becomes independent on the starting point $\bf{x}_{0}^{0}$ , $%
\bf{C}_{0}^{0}$ as k is increased.

When the last plane (nth) is taken into account the Kalman filter
performs the final step which is a smoothing. The filter runs
backward in time updating all filtered state vectors on the basis
of information from all n planes. The equations describing
\textbf{the smoothing} are given by

\[
\bf{A}_{k}=\bf{C}_{k}^{k}\bf{F}_{k}^{T}(\bf{C}_{k+1}^{k})^{-1}
\]

\[
\bf{x}_{k}^{n}=\bf{x}_{k}^{k}+\bf{A}_{k}(\bf{x}_{k+1}^{n}-\bf{x}_{k+1}^{k})
\]

\[
\bf{C}_{k}^{n}=\bf{C}_{k}^{k}+\bf{A}_{k}(\bf{C}_{k+1}^{n}-\bf{C}_{k+1}^{k})\bf{A}_{k}^{T}
\]

\[
\bf{r}_{k}^{n}=\bf{m}_{k}-\bf{H}_{k}\bf{x}_{k}^{n}
\]

\[
\bf{R}_{k}^{n}=\bf{V}_{k}-\bf{H}_{k}\bf{C}_{k}^{n}\bf{H}_{k}^{T}
\]

   Until now it was assumed that the problem of estimation of a discrete-time
process is described by a linear stochastic differential equation.
However for example in the presence of a magnetic field the track
propagator $\bf{F}$ is non-linear.
   Let's assume that the process of a particle propagation is governed by the
non-linear stochastic differential equation

\begin{equation}
x_{k}=f(x_{k-1})+w_{k-1}  \label{8ap}
\end{equation}

with a measurement m in the form Eq.(\ref{2ap}). f is a non-linear
function. The Kalman filter can be applied to this system by
linearizing the system for example about the estimated trajectory.
If deviations between the estimated trajectory and the actual
trajectory remain sufficiently small the linear approximation is
valid.
   The non-linear equation (\ref{8ap}) can be written down in the linearized form
as

\begin{equation}
\bf{x}_{k}=\bf{f}(\bf{x}_{k-1}^{k-1})+\bf{F}
\cdot(\bf{x}_{k-1}-\bf{x}_{k-1}^{k-1})+\bf{w}_{k-1}\label{10ap}
\end{equation}

where as before $\bf{x}_{k}$,$\bf{m}_{k}$ are the actual
state and measurement vectors, $\bf{x}_{k}^{k}$ is a filtered
estimate of the state at step k. $\bf{F}$ is Jacobian matrix

\begin{equation}
\bf{F}_{ij}=\partial
\bf{f}_{i}(\bf{x}_{k-1}^{k-1})/\partial x_{j} \label{11ap}
\end{equation}

Therefore the complete set of extended Kalman filter equations is
given by Eqs.(\ref{4ap})-(\ref{7ap}),(\ref{10ap}) by using F in
the form (\ref{11ap}).

In order to apply the extended Kalman filter to a track fitting
for a particle moving in uniform magnetic field (the magnetic
field is in z direction) one has to choose the state vector
parameters, define the initial state vector and calculate the
transport matrix $\bf{F}$, the projection matrix $\bf{H}$, and
the noise matrix $\bf{Q}$.
   As it was mentioned above in this case the state vector can be chosen
in the form ${\bf{x}_{k}}=(x,y,t_{x},t_{y},1/p_{L})$ where x, y
are the track coordinates in the tracker system,
$t_{x}=p_{x}/p_{L}$, $t_{y}=p_{y}/p_{L}$ define the track
direction.
   The projection matrix H is given by Eq.(\ref{2ap1}).
   Due to multiple scattering the absolute value of electron momentum
remains unaffected, while the direction is changed. This
deflection can be described using two orthogonal scattering
angles, which are also orthogonal to the particle momentum ~\cite{mank}.
 In terms of these variables the noise
matrix is given by

$$
\bf{Q}_{k}=<\Theta ^{2}>(t_{x}^{2}+t_{y}^{2}+1) \left.\left(
\begin{array}{ccccc}
0 & 0 & 0 & 0 & 0 \\
0 & 0 & 0 & 0 & 0 \\
0 & 0 & t_{x}^{2}+1 & t_{x}t_{y} & t_{x}/p_{L} \\
0 & 0 & t_{x}t_{y} & t_{y}^{2}+1 & t_{y}/p_{L} \\
0 & 0 & t_{x}/p_{L} & t_{y}/p_{L} & \frac{(t_{x}^{2}+t_{y}^{2})}{
p_{L}^{2}(t_{x}^{2}+t_{y}^{2}+1)}
\end{array}
\right.\right)$$

   For the variance of the multiple scattering angle the well-known
expression is used

\begin{equation}
<\Theta^{2}>=(13.6MeV/p)^{2}[1+0.038\ln(t/X_{R})]t/X_{R}
\end{equation}

where $X_{R}$ is a radiation length, t is a distance traveled by
the particle inside a scatterer.
   Energy losses are taken into account by

\begin{equation}
p\prime = p - <dE/dx>t.
\end{equation}

\qquad

\section*{APPENDIX\ B. DETERMINISTIC ANNEALING FILTER}

Track reconstruction in modern high energy physics 
experiments faces a significant
amount of noise hits in a detector. The track fit thus is 
confronted with several competing hits in detector's layers.
The Kalman filter (see Appendix A) is 
now widely used
for the reconstruction of the track parameters in high energy
physics. However the 
application of the Kalman filter requires that the problem 
of assignment of the detector hits to track candidates has been
entirely resolved by the preceding selection procedure. 
If this is not the case, the filter has to run on every 
possible assignment to select the best one by chi-square
criterion. Obviously this approach is computationally expensive
and practically unfeasible for a considerable amount of noise
hits. For this reason the Deterministic Annealing Filter (DAF)
was developed ~\cite{daf}. In DAF there is an additional 
validation feature eliminating hits which are not compatible
with the predicted track position.

The deterministic annealing filter itself is a Kalman filter
with re-weighted observations. The propagation part of DAF is
identical to the standard Kalman filter. 

\textbf{The filtered estimate} (measurement update)
$\bf{x}_{k}^{k}$ at layer k is calculated as a weighted mean of the
prediction $\bf{x}_{k}^{k-1}$ and the observations 
{$\bf m_{k}^{i}, i=1,2,...n_{k}$}:

\begin{equation}
\bf x_{k}^{k}=\bf x_{k}^{k-1}+\bf K_{k}\sum _{i=1}^{{n}_{k}} 
p_{k}^{i}[\bf{m}_{k}^{i}-\bf{H}_{k}\bf{x}_{k}^{k-1}]
\label{1bp}
\end{equation}

where ${\bf p_{k}^{i}}$ is the assignment probability of observation $\bf m_{k}^{i}$.
$\bf{K}_{k}$ is the Kalman gain matrix which is given by  

\begin{equation}
\bf{K}_{k}=[[\bf{C}_{k}^{k-1}]^{-1}+p_{k}\bf{H}_{k}^{T}\bf{V}_{k}^{-1}\bf{H}_{k}]^{-1}
\bf{H}_{k}^{T}\bf{V}_{k}^{-1}
\label{2bp}
\end{equation}

where ${\bf p_{k}}$ is the sum over all weights ${\bf p_{k}^{i}}$, $\bf{H}_{k}$ is the 
measurement matrix, $\bf{V}_{k}$ is the variance of the observations.

The covariance matrix $\bf{C}_{k}^{k}$ of the updated estimate
$\bf{x}_{k}^{k}$ is written as 

\begin{equation}
\bf{C}_{k}^{k}=[[\bf{C}_{k}^{k-1}]^{-1}+p_{k}\bf{H}_{k}^{T}\bf{V}_{k}^{-1}\bf{H}_{k}]^{-1}.
\label{3bp}
\end{equation}
  
After completion of the forward filter a backward filter runs
in opposite direction, using the same weights as the forward
filter. By taking a weighted mean of the filtered states of 
both filters at every layer a prediction for the state vector 
$\bf x_{k}^{n*}$ along with its covariance matrix $\bf {C}_{k}^{n*}$ 
is obtained, using all hits except the ones at layer k. (The asterisk
indicates that the information from layer k is not used in 
this prediction.) Initially all assignment probabilities for the
hits in each layer 
are set
to be equal but based on the estimated state vector 
$\bf x_{k}^{n*}$ and its covariance 
matrix, the assignment probabilities of all competing hits are then 
recalculated in the following way:

\begin{equation}
{\bf {p}_{k}^{i}} \sim \varphi(\bf{m}_{k}^{i};
\bf{H}_{k}\bf x_{k}^{n*}, \bf{V}_{k}+\bf{H}_{k}\bf{C}_{k}^{n*}\bf{H}_{k}^{T})
\label{4bp}
\end{equation}

where $\varphi(\bf x ; {\bf \mu},\bf V$) is a multivariate Gaussian 
probability density with mean vector ${\bf \mu}$ and covariance
matrix $\bf V$. 

If the probability falls below a certain threshold, the hit is 
considered as the false one and is excluded from the list of the
hits assigned to the track. 

However at this step we cannot be sure in calculated probabilities
especially in the initial phase due to insufficient information for
the filter. This problem is overcome by adopting a simulated annealing     
iterative procedure. This is an additional feature of DAF.

The simulated annealing optimization algorithm is based on an
analogy between the behavior of a material heated past its
melting point that is slowly cooled (annealed) to form a single
crystal. If the cooling proceeds slowly enough, the crystalline 
state reached at zero temperature will have all the atoms fixed 
in a perfect lattice structure, corresponding to the lowest 
possible energy of the system (global minimum).

In the same way in track fitting the simulated annealing allows
to avoid a local minimum and find the global one corresponding to
the minimum chi-square for the track.

DAF annealing algorithm can be described in the following way.
The annealing schedule is chosen for example in the form
${\bf V_{N}}={\bf V}(A/f^{N}+1)$ where the annealing factor 
$f>1$ and factor $A>>1$. This provides that the initial 
variance  is well above the nominal value 
${\bf V}$ of the observation error but the final one tends 
to ${\bf V}$. After each iteration the assignment 
probabilities exceeding the threshold are normalized to 1
and used again as weights in the next iteration, and so on.
The iterations generally are stopped if the relative change in 
chi-square is less than correspondent control parameter
(typically of the order 0.01). 

Since we deal with the stochastic process the best result
can be reached repeating the DAF procedure for a few different
annealing factors f and then choosing the result 
corresponding to the minimum chi-square.

\newpage

\section*{APPENDIX\ C. Tracker Resolution}

This appendix demonstrates the results of application of the pattern
recognition and reconstruction procedure for the conversion events 
without background.

\begin{figure}[htb!]
\centerline{\hbox{%
\psfig{figure=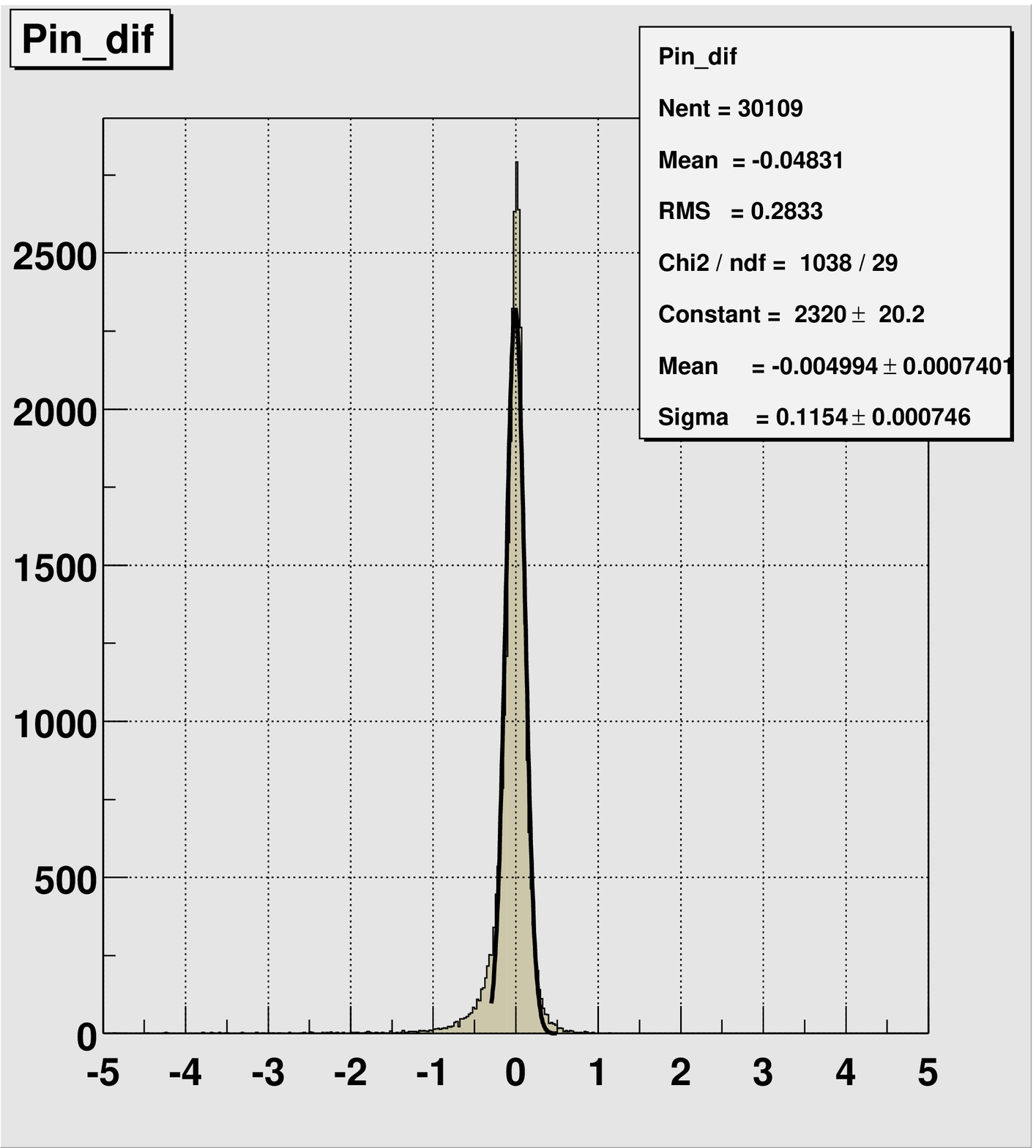,height=3.0in,clip=on}
 \psfig{figure=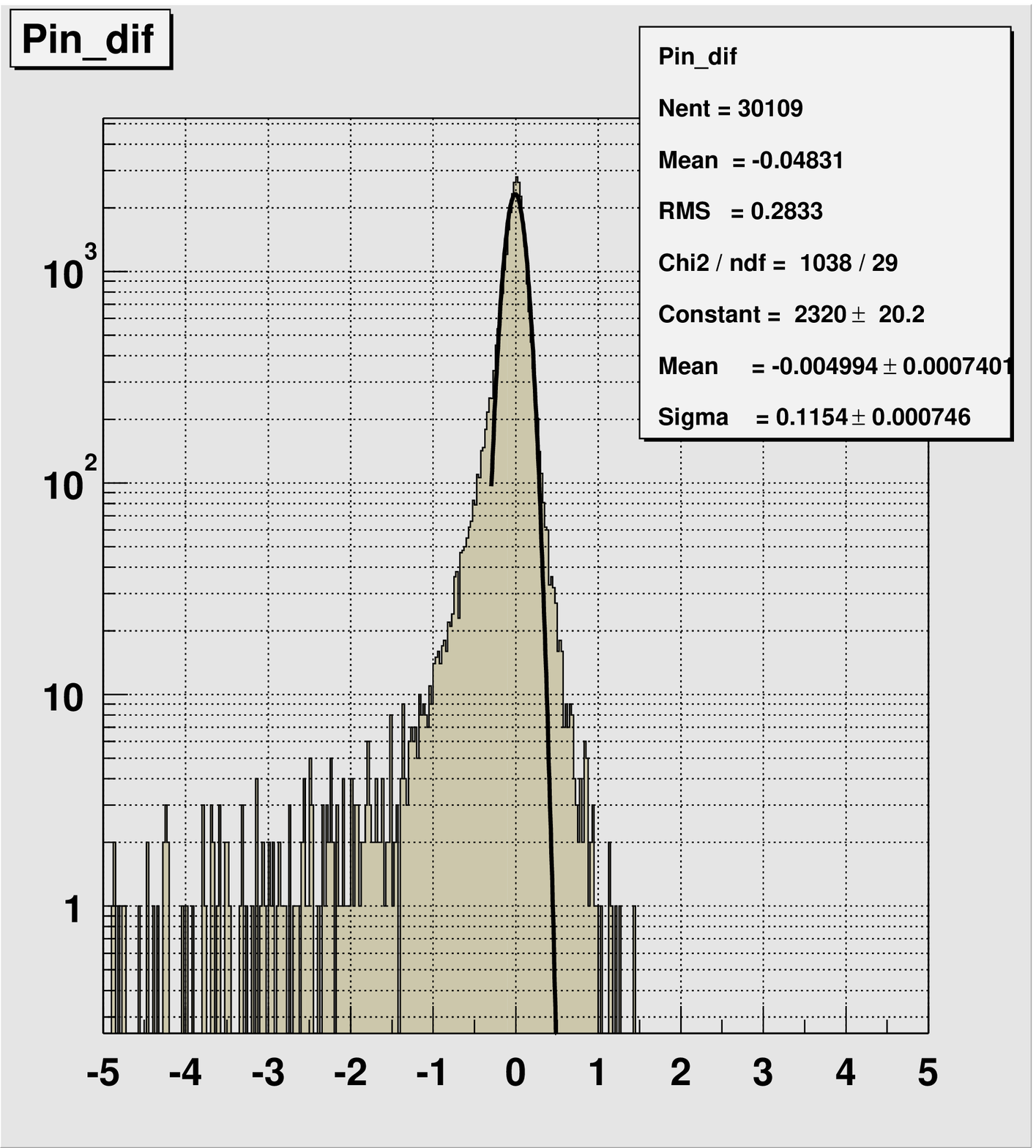,height=3.0in,clip=on}}}
\caption{
 Distribution in the difference between the input reconstructed momentum based on
 the Kalman filter and the simulated input momentum without background.
 }
\label{fig:pin_dif}
\end{figure}

 The distribution in the difference between the input reconstructed momentum (Pin\_f) based on
 the Kalman filter and the simulated input momentum (Pin) is shown in
Figure ~\ref{fig:pin_dif} in linear (a) and logarithmic (b) scale.
According to this distribution the intrinsic tracker resolution is
$\sigma$ = 0.12 MeV if one fits the distribution by a Gaussian in the range
-0.3 - 0.7 MeV.

\begin{figure}[htb!]
\centerline{\hbox{%
\psfig{figure=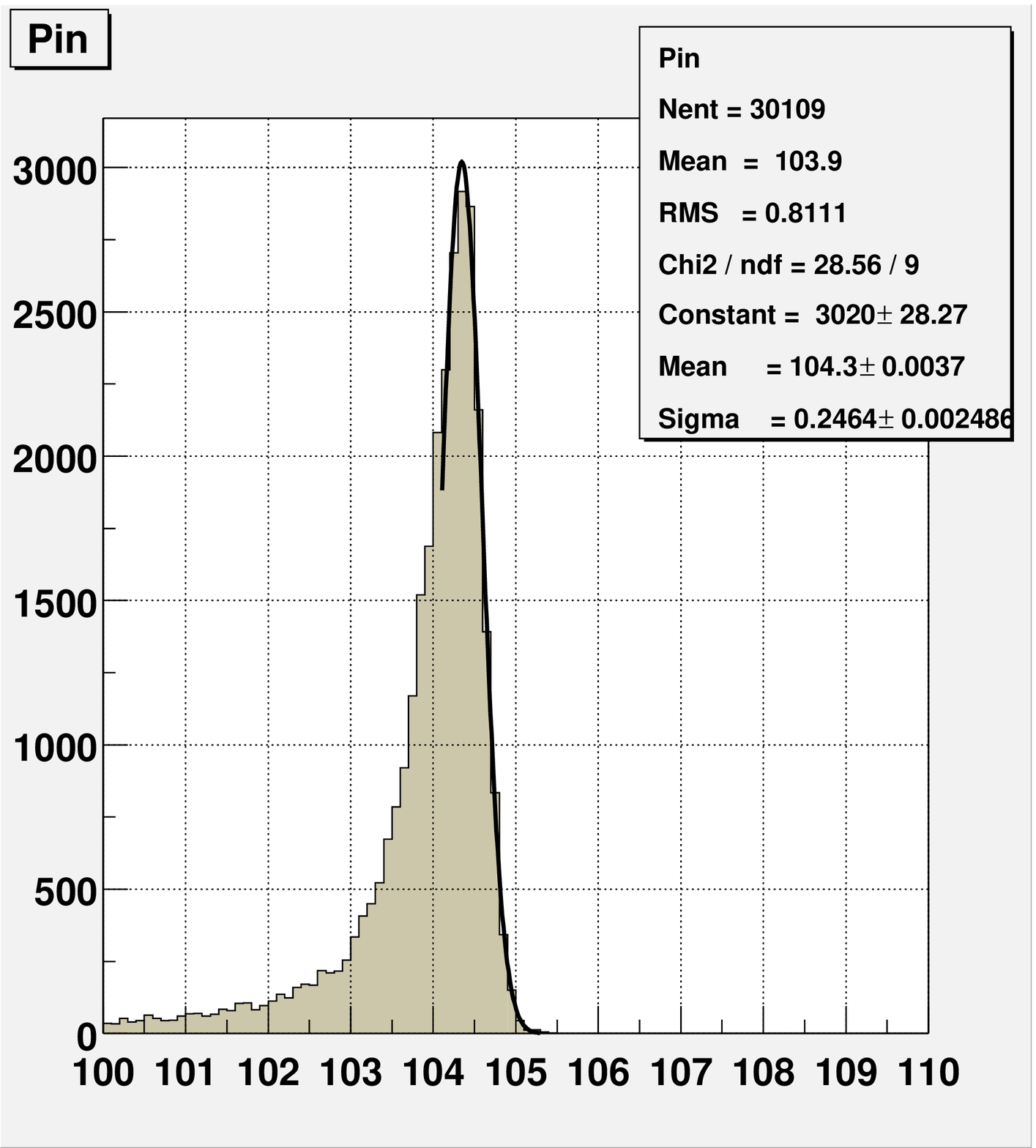,height=3.0in,clip=on}
 \psfig{figure=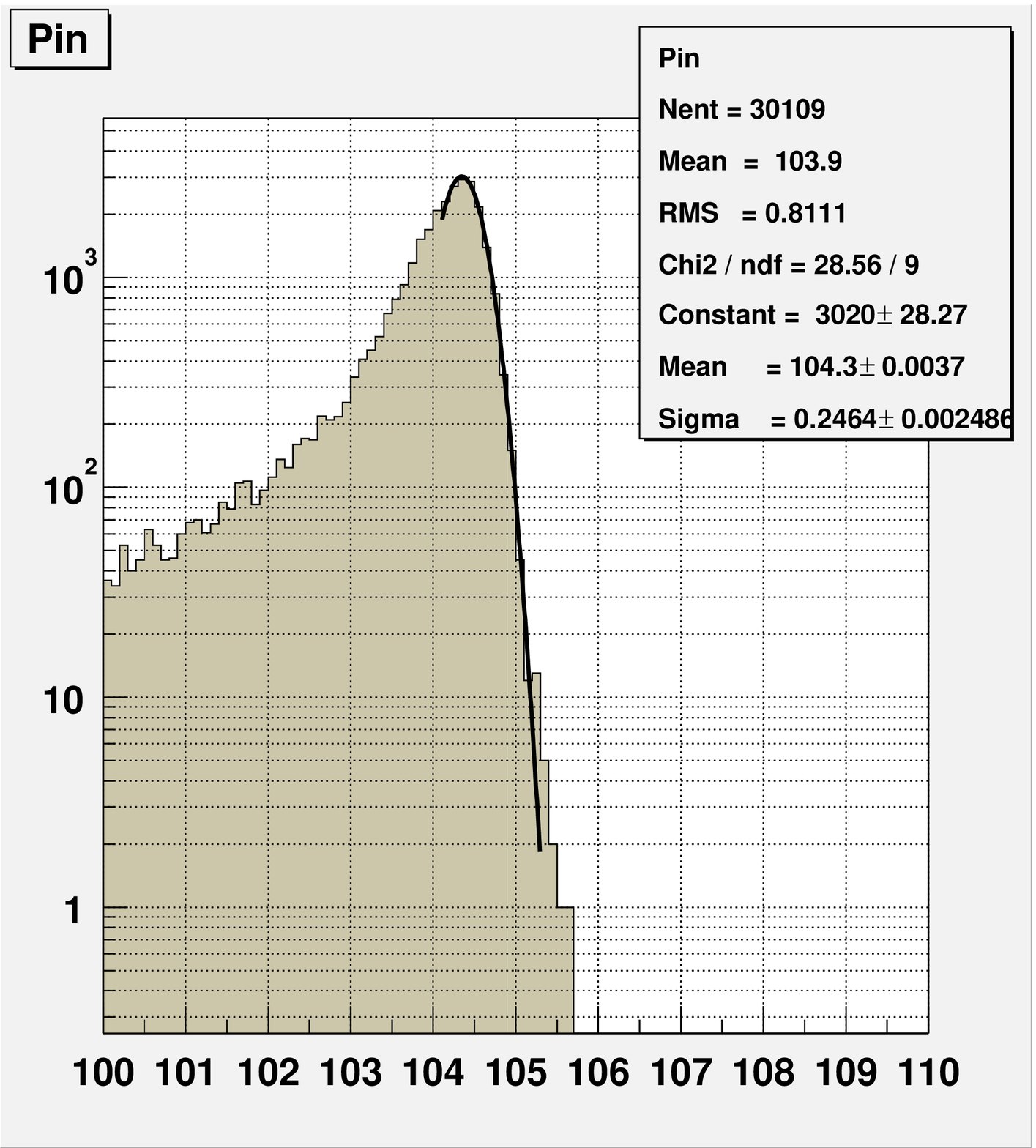,height=3.0in,clip=on}}}
\caption{
 Distribution in the input momentum reconstructed by the
 Kalman filter  without background.
 }
\label{fig:pin_kal}
\end{figure}

Figure ~\ref{fig:pin_kal} shows a distribution in the input
momentum (Pin\_f) reconstructed by the Kalman filter 
in linear (a) and logarithmic (b) scale without background.  
This distribution is characterized by the standard deviation $\sigma$ = 0.25 MeV of the
reconstructed input momentum  for a Gaussian fit
in the range 104-106 MeV. 
The overall reconstruction acceptance is 22.7 $\%$ for muon conversion
events with the momentum above a threshold momentum of 103.6
MeV/c.

Comparing these results with the results of the reconstruction
in the presence of the background we get the difference in tracker
resolution 1.5 \% and the difference 2.7 \% in overall acceptance.
Therefore the tracker resolution and overall acceptance are not
affected significantly at the considered background level.

\end{document}